\documentclass[%
 aip,
 amsmath,amssymb,
 reprint,%
]{revtex4-1}

\usepackage{graphicx}
\usepackage{dcolumn}
\usepackage{bm}

\usepackage[utf8]{inputenc}
\usepackage[T1]{fontenc}
\usepackage{etoolbox}

\usepackage{mathtools,amsmath}
\usepackage{graphicx} 
\usepackage[breaklinks=true,colorlinks,citecolor=teal,linkcolor=teal,urlcolor=teal]{hyperref}
\usepackage{physics}
\usepackage{siunitx}
\usepackage{bbold}
\usepackage{soul}
\usepackage{bm}
\usepackage[normalem]{ulem}
\usepackage{times}
\usepackage{titlesec}
\usepackage{enumerate}  
\usepackage{float}
\usepackage{amssymb}
\usepackage{algorithm2e}
\usepackage{newtxmath}
\RestyleAlgo{ruled}
\SetKwComment{Comment}{/* }{ */}
\usepackage{xcolor}
\usepackage[title]{appendix}

\newcommand{\mcs}{Department of Mathematics and Computer Science, Freie Universit{\" a}t Berlin, Arnimallee 6, 14195 Berlin, Germany}
\newcommand{\dccqs}{Dahlem Center for Complex Quantum Systems, Freie Universit{\"a}t Berlin, Arnimallee 14, 14195 Berlin, Germany}
\newcommand{\msr}{Microsoft Research AI4Science, Karl-Liebknecht Str. 32, Berlin 10178, Germany}
\newcommand{\ph}{Department of Physics, Freie Universit{\" a}t Berlin, Arnimallee 14, 14195 Berlin, Germany}
\newcommand{\rice}{Department of Chemistry, Rice University, 6100 Main St, Houston, Texas 77005, USA}
\newcommand{\cnrs}{JEIP, USR 3573 CNRS, Coll{\` e}ge de France, PSL Research University, 11 Place Marcelin Berthelot, F-75321 Paris, France}

\makeatletter
\def\@email#1#2{%
 \endgroup
 \patchcmd{\titleblock@produce}
  {\frontmatter@RRAPformat}
  {\frontmatter@RRAPformat{\produce@RRAP{*#1\href{mailto:#2}{#2}}}\frontmatter@RRAPformat}
  {}{}
}%
\makeatother
\begin{document}

\preprint{AIP/123-QED}

\title[Deep quantum Monte Carlo approach for polaritonic chemistry]{Deep quantum Monte Carlo approach for polaritonic chemistry}
\author{Yifan Tang}
\homepage{yifta@zedat.fu-berlin.de}
\affiliation{\mcs} \affiliation{\dccqs}

\author{Gian Marcello Andolina}
\affiliation{\cnrs}

\author{Alice Cuzzocrea}
\affiliation{\mcs}

\author{Mat{\v e}j Mezera}
\affiliation{\mcs}

\author{P. Bern{\' a}t Szab{\' o}}
\affiliation{\mcs}

\author{Zeno Sch{\" a}tzle}
\affiliation{\mcs}

\author{Frank No{\' e}}
\affiliation{\msr} \affiliation{\mcs} \affiliation{\ph} \affiliation{\rice}

\author{Paolo A. Erdman}
\homepage{p.erdman@fu-berlin.de}
\affiliation{\mcs}

\date{\today}

\begin{abstract}
Recent years have witnessed a surge of experimental and theoretical interest in controlling the properties of matter, such as its chemical reactivity, by confining it in optical cavities, where the enhancement of the light-matter coupling strength leads to the creation of hybrid light-matter states known as polaritons. However, ab initio calculations that account for the quantum nature of both the electromagnetic field and matter are challenging and have only started to be developed in recent years. We introduce a deep learning variational quantum Monte Carlo method to solve the electronic and photonic Schr\"odinger equation of molecules trapped in optical cavities. We extend typical electronic neural network wavefunction ansatzes to describe joint fermionic and bosonic systems, i.e. electron-photon systems, in a quantum Monte Carlo framework. We apply our method to hydrogen molecules in a cavity, computing both ground and excited states. We assess their energy, dipole moment, charge density shift due to the cavity, the state of the photonic field, and the entanglement developed between the electrons and photons. When possible, we compare our results with more conventional quantum chemistry methods proposed in the literature, finding good qualitative agreement, thus extending the range of scientific problems that can be tackled using machine learning techniques. 
\end{abstract}

\maketitle

\section{Introduction}
Recently, there has been a growing interest in understanding the impact of strong light-matter coupling on the physical and chemical properties of matter~\cite{Byrnes2014Exciton,Hagenmueller2017Cavity,Wang2019Cavity,Garcia2021Manipulating, Schlawin2022Cavity, Bloch2022Strongly}. For example, when a molecular system is placed in an optical cavity, the interaction between the system and the quanta of the electromagnetic field gives rise to hybrid light-matter states~\cite{Siegman2000Laser,Herrera2016Cavity, Haroche2013Nobel, Bernardis2024Light}, known as polaritons, which may display profoundly different physical and chemical properties~\cite{Basov2016Polaritons,Wang2021Light,Fregoni2022Theoretical, Sidler2022A, Sidler2023Numerically, Sidler21JPCL, Ruggenthaler2023Understanding, Ebbesen2023Introduction, Sokolovskii2023Multi, Barquilla2022A}. Their influence on the molecular properties of matter can be modulated by tuning the cavity parameters. With the improvement of optical cavity engineering, devices can now reach the strong coupling regime, where the rate of energy exchange between light and matter is large compared to the photon leakage rate and non-radiative losses of the emitter~\cite{Frisk2019Ultrastrong,Le2020Theoretical, Lednev2024Lindblad, Jarc2023Cavity, Appugliese2022Breakdown}, even at room temperature~\cite{Chikkaraddy2016Single, Bujalance2024Strong}. In this regime, the appearance of polaritons can have profound effects on molecular systems, such as modifying reaction rates~\cite{Thomas2016Ground,Galego2019Cavity,Thomas2019Tilting} and changing the chemical environment~\cite{Castagnola2024Collective}.

From a theoretical perspective, understanding these effects requires a proper description of the strong electron-photon coupling, which could serve as a guide in designing polaritonic chemical experiments. Polaritonic chemistry is deeply intertwined with quantum effects, as some phenomena are persistent even in the absence of illumination due to the interaction with quantum vacuum fluctuations. Therefore, a fully quantum treatment is necessary, including the description of electromagnetic fields using quantum electrodynamics (QED)~\cite{Jaynes1963Comparison,Cumming1965Stimulated, Svendsen2024Ab}. Recently, several ab initio methods have been developed, including QED Hartree-Fock~\cite{Haugland2020Coupled,Riso2022Molecular}, quantum-electrodynamical density-functional theory (QEDFT)~\cite{Ruggenthaler2014Quantum,Flick2017Atoms,Foley2023}, QED coupled cluster theory (QED-CC)~\cite{Haugland2020Coupled, Pavosevic2021Polaritonic, Castagnola2024Strong}, and QED full configuration interactions (QED-FCI)~\cite{Haugland2020Coupled, Haugland2021Intermolecular}. QEDFT, an extension of the density functional theory (DFT), can be applied to molecules on a large scale. However, it suffers from limited accuracy due to incomplete information on the exchange-correlation functional in the electron-photon interaction. The coupled cluster (CC) and configuration interaction (CI) methods can reach a satisfactory accuracy, but their computational costs increase sharply with the system size. Indeed, for electronic systems, the cost scales as $\mathcal{O}(N^6)$ for CC with single and double excitations (CCSD)~\cite{Bartlett2007Coupled} and $\mathcal{O}(\exp{N})$ for FCI with $N$ orbitals~\cite{szabo1996modern}. In these approaches, the electronic equations are solved in the basis of Slater determinants, and their number is typically large because of the difficulty in accurately describing the ground state of quantum many-body systems~\cite{Shavitt2009Many}. In general, all methods present an inevitable tradeoff between computational cost and accuracy.

In the past decade, machine learning has proven to be a powerful tool in diverse fields, ranging from image and language processing~\cite{LeCun1989Handwritten, LeCun2015Deep}, to scientific applications such as computational physics and chemistry~\cite{Carleo2019Machine,Carrasquilla2021How,Keith2021Combining} . One approach is to use supervised learning which, utilizing huge amounts of data and state-of-the-art algorithms, takes a detour from solving the Schr\"odinger equations directly and attempts to design a model to predict chemical properties of molecules~\cite{Chmiela2017Machine,Smith2017ANI-1,Schutt2018SchNet,Schutt2019Unifying}. While this approach has a great advantage in terms of computational cost, its accuracy is largely dependent on the quality of the datasets, which could be expensive to produce, may have limited accuracy, and typically  requires domain knowledge and expertise to produce. 

Another approach is given by deep-learning quantum Monte Carlo (deep QMC), an ab-initio technique that does not require prerequisite knowledge of the solutions. First proposed by Carleo and Troyer and applied to spin-lattice systems in Ref.~\cite{Carleo2017Solving}, deep QMC represents wavefunctions with deep neural networks (DNN) and trains them based on the variational principle~\cite{Hermann2023Ab}. The accuracy of the QMC method is fundamentally determined by the wavefunction ansatz, which is usually efficiently represented by a DNN. Such DNNs encode electronic properties of the system in its structure; the high expressiveness of the ansatz is then ensured by the large amount of tunable parameters. These parameters are optimized during the training process so that the corresponding wavefunction represents a good approximation to the exact eigenstates of the Hamiltonian. Deep QMC has been applied to the study of the electronic structure of molecules~\cite{Han2019Solving, Hermann2020Deep, Pfau2020Ab, vonGlehn2023A}. It proves successful in the calculation of ground and excited state energies~\cite{Entwistle2023Electronic, Szabo2024An, pfau2024accurate}, in studying periodic systems~\cite{Pescia2022Neural}, and in calculating interatomic forces~\cite{Qian2022Interatomic}.

In this work, we introduce a deep QMC approach to study the ground and excited states of molecules in a cavity. We propose a novel approach to promote electronic deep neural network wavefunction ansatzes to electron-photon ansatzes that can describe both the electronic and photonic degrees of freedom of the system and, thus, electron-photon correlations. Our extension, which allows to describe joint fermionic and bosonic systems, can be applied to many electronic wavefunction ansatzes including PauliNet~\cite{Hermann2020Deep} and its variants~\cite{Schaetzle2021Convergence, Entwistle2023Electronic}. Here, as an example, we use the PauliNet2 ansatz, introduced and implemented in Ref.~\cite{Schaetzle2023Deep}. Our method enjoys both the high accuracy of QMC methods and the fast speed achieved by parallel computation with GPUs. We implement our method within the \texttt{DeepQMC} package~\cite{hermann2024deepqmc} based on the JAX library~\cite{JAX2018Github}, including our extension of PauliNet2 to describe molecules in cavities, the corresponding light-matter Hamiltonian, and a new sampling method to account for both electronic and photonic degrees of freedom which are treated in first and second quantization respectively.

The paper is organized as follows. We present the deep QMC method for electron-photon systems in Sec.~\ref{sec:photon_deepQMC}. Here we introduce the Hamiltonian describing the light-matter interaction, we frame the calculation of the ground and excited states as a machine-learning problem, we introduce our wavefunction ansatz extension to electronic and photonic degrees of freedom, and we show how to compute the loss function and physical observables in a quantum Monte Carlo framework. In Sec.~\ref{sec:results} we apply our methods to small molecular systems, investigating both the ground and excited states of the system, and we discuss how various system observables are modified by the cavity. Sec.~\ref{sec:conclusions} summarizes our results and provides an outlook for future research. All equations are expressed in atomic units, i.e., $e=m_{\text{e}}=\hbar=c=1/(4\pi\epsilon_0)=1$.

\section{Deep quantum Monte Carlo for electron-photon systems}\label{sec:photon_deepQMC}
In this section, we propose a deep QMC method to solve the Schr\"odinger equation of molecules confined in an optical cavity. We first introduce
the electron-photon coupling model in Subsec.~\ref{subsec:Hamiltonian}, given by the Pauli-Fierz Hamiltonian~\cite{Spohn2004Dynamics, Rokaj2018Light}. In Subsec.~\ref{subsec:vm_loss} we define the loss function for the machine learning task using the variational principle. In Subsec.~\ref{subsec:el-ph_ansatz} we propose a neural network wavefunction extension that describes the electronic degrees of freedom in first quantization, and the photonic degrees of freedom in second quantization, allowing us to describe correlations and entanglement between them. The Monte Carlo method for evaluating arbitrary physical observables, including the system energy and its gradient, is presented in Subsec.~\ref{subsec:obs_el} and a sampling algorithm dealing with both discrete and continuous variables is introduced in Subsec.~\ref{subsec:dc_algo}.
Finally, we summarize the method with a flowchart of the whole procedure in Subsec.~\ref{subsec:summary}.

\subsection{Electron-photon coupling model}\label{subsec:Hamiltonian}
We consider a molecular system with $N_{\text{e}}$ electrons and $N_{\text{nuc}}$ nuclei in an optical cavity. For simplicity, we only consider a single cavity mode corresponding to the one with the strongest effect, although a generalization of our method to the multi-mode case is conceptually straightforward. We denote the frequency of the mode as $\omega$, the coupling strength as $\lambda=\sqrt{\frac{4\pi}{V}}$, where $V$ is the quantization volume of the electromagnetic field, and the polarization vector as $\boldsymbol{\varepsilon}$. 
The Pauli-Fierz Hamiltonian of the light-matter interacting system is derived in the dipole gauge using a series of customary approximations~\cite{Schaefer2020Relevance,Haugland2020Coupled, Haugland2021Intermolecular, Riso2022Molecular}, consisting of the Born-Oppenheimer approximation, the dipole approximation~\cite{Rokaj2018Light}, and the PZW transformation~\cite{Woolley2020PZW}, resulting in the following form:
\begin{equation}\label{eq:Hamiltonian_all}
H_\text{PF}
=H_\text{e}+\omega \hat{b}^\dagger \hat{b}
-\sqrt{\frac{\omega}2}\lambda(\boldsymbol{\varepsilon}\cdot\mathbf{d})(\hat{b}+\hat{b}^\dagger)
+\frac12\lambda^2(\boldsymbol{\varepsilon}\cdot\mathbf{d})^2,
\end{equation}
\noindent with molecular dipole operator 
\begin{equation}
\mathbf{d}=-\sum\limits_{i=1}^{N_\text{e}}\mathbf{r}_i+\sum\limits_{I=1}^{N_\text{nuc}}Z_I\mathbf{R}_I
\end{equation}
for electron positions $\{\mathbf{r}_i\}_{i=1}^{N_\text{e}}$ and nuclear positions $\{\mathbf{R}_I\}_{I=1}^{N_{\text{nuc}}}$.
The first term $H_\text{e}$ in Eq.~\eqref{eq:Hamiltonian_all} represents the standard electronic Hamiltonian, given by
\begin{eqnarray}\label{eq:Hamiltonian_electronic}
H_{\text{e}}&=&-
\sum\limits_{i=1}^{N_\text{e}}\frac12\nabla_i^2
-\sum\limits_{i=1}^{N_\text{e}}\sum\limits_{I=1}^{N_{\text{nuc}}}\frac{Z_I}{|\mathbf{r}_i-\mathbf{R}_I|}\nonumber
 \\ 
&&+ \frac12\sum\limits_{i\neq j}^{N_\text{e}}\frac1{|\mathbf{r}_i-\mathbf{r}_j|}+ 
\frac12\sum\limits_{I\neq J}^{N_{\text{nuc}}}\frac{Z_IZ_J}{|\mathbf{R}_I-\mathbf{R}_J|}.
\end{eqnarray}
The second term $\omega \hat{b}^\dagger \hat{b}$ represents the Hamiltonian of the cavity mode, where $\hat{b}$ and $\hat{b}^\dagger$ are respectively the photonic annihilation and creation operators. The following terms describe the electron-photon interaction. 

\subsection{Variational method and the loss function}\label{subsec:vm_loss}
We wish to solve the time-independent electronic Schr\"odinger equation
\begin{equation}\label{eq:Schrodinger}
H_\text{PF}|\psi\rangle=E|\psi\rangle
\end{equation}
\noindent for the eigenenergies and corresponding eigenstates of $H_\text{PF}$. Thanks to the variational principle, the ground state energy $E_0$ can be obtained by solving the minimization problem
\begin{equation}\label{eq:optimization}
E_0 \leq \min\limits_\theta E[\psi_\theta]=\min\limits_\theta\frac{\langle\psi_\theta|H_\text{PF}|\psi_\theta\rangle}{\langle\psi_\theta|\psi_\theta\rangle},
\end{equation}
\noindent where the trial wavefunction $\ket{\psi_\theta}$ (the ansatz) is parametrized by a set of parameters $\theta$. In the limit of an infinitely expressive ansatz, the inequality in Eq.~(\ref{eq:optimization}) becomes a strict equality.
To state this as a machine learning problem, we choose the energy in Eq.~\eqref{eq:optimization} as the loss function $\mathcal{L}(\theta)$. To solve for excited states, penalty terms have been introduced to ensure that different eigenstates are orthogonal and have the desired spin value. Introducing an ansatz for the ground state $\ket*{\psi^{(0)}_\theta}$ and for the lowest $N_\text{s}$ excited states $\left\{\ket*{\psi^{(1)}_\theta},\cdots,\ket*{\psi^{(N_\text{s})}_\theta}\right\}$, we consider the loss function~\cite{Szabo2024An} 
\begin{eqnarray}
\label{eq:loss_function_ext}
\mathcal{L}(\theta)&=&\sum\limits_{k=0}^{N_\text{s}}\frac{\langle\psi^{(k)}_\theta|H_\text{PF}+\beta S^2|\psi^{(k)}_\theta\rangle}{\langle\psi^{(k)}_\theta|\psi^{(k)}_\theta\rangle}\nonumber \\
&&+ \sum\limits_{k<l}\alpha_{k,l}\frac{|\langle\psi^{(k)}_\theta|\psi^{(l)}_\theta\rangle|^2}{\langle\psi^{(k)}_\theta|\psi^{(k)}_\theta\rangle\langle\psi^{(l)}_\theta|\psi^{(l)}_\theta\rangle}.
\end{eqnarray}
\noindent The first sum in Eq.~\eqref{eq:loss_function_ext} represents the total energy of the ground and excited states, including an additional penalty term $\beta S^2$ used to enforce the desired spin value,
where $S^2$ is the squared spin operator, and $\beta>0$ is a hyperparameter \cite{Szabo2024An}.
The second sum is a penalty term enforcing the orthogonality between the states, where $\{\alpha_{k,l}\}$ is a set of hyperparameters satisfying $\alpha_{k,l} > E_l-E_k$ for eigenenergies $E_k$ and $E_l$~\cite{liu2023calculate, Entwistle2023Electronic}. By minimizing Eq.~\eqref{eq:loss_function_ext} we can find an approximation to the ground state and to the first $N_\text{s}$ excited states within the desired spin sector.

\subsection{Electron-photon wavefunction ansatz}\label{subsec:el-ph_ansatz}
Here we propose a wavefunction ansatz to describe the eigenstates of molecules in an optical cavity, which is a joint fermionic and bosonic system. The idea is to extend common electronic ansatzes, that describe electrons in first quantization, with a second quantization description of the photonic state.
The Fock state basis of a single-mode field is $\{|n\rangle\mid n\in\mathbb{N}\}$, consisting of the eigenstates of the Hamiltonian $\omega \hat{b}^\dagger \hat{b}$ corresponding to a harmonic oscillator with frequency $\omega$~\cite{Fock1932Konfigurationsraum}. We have $\hat{b}^\dagger|n\rangle=\sqrt{n+1}|n+1\rangle, \hat{b}|n\rangle=\sqrt{n}|n-1\rangle.$ We consider the basis for the electron-photon wavefunction as the tensor product of the electronic positional basis $\{|\mathbf{r}\rangle\}$, where $\mathbf{r}=(\mathbf{r}_1,\cdots,\mathbf{r}_{N_\text{e}})\in\mathbb{R}^{3N_\text{e}}$ denotes the positions of all electrons,
and the Fock state basis $\{|n\rangle\}$:
\begin{equation}\label{eq:electron-photon_basis}
|\mathbf{r},n\rangle=|\mathbf{r}\rangle\otimes|n\rangle.
\end{equation}
\noindent It is a complete basis since both $\{|\mathbf{r}\rangle\}$ and $\{|n\rangle\}$ are complete. Expanding the electron-photon wavefunction $|\psi\rangle$ in this basis
\begin{equation}\label{eq:electron-photon_expansion_ph}
|\psi\rangle=\sum\limits_{n=0}^\infty\int\mathrm{d}\mathbf{r}|\mathbf{r},n\rangle\langle\mathbf{r},n|\psi\rangle,
\end{equation}
\noindent we are interested in parameterizing the coefficients $\psi(\mathbf{r},n)=\langle\mathbf{r},n|\psi\rangle$ using deep neural networks. In practice, we truncate the summation in Eq.~\eqref{eq:electron-photon_expansion_ph} to $n\in\{0,1,\cdots,N_{\text{max}}\}$, where $N_\text{max}$ is a cutoff value, since the coefficients for the lowest values of $n$ are dominant when studying the ground state and low-lying excited states. 
A DNN representation of the electron-photon wavefunction ansatz $\psi_\theta(\mathbf{r},n)$ can be obtained from common DNN electronic wavefunction ansatzes adding an additional discrete variable $n$ to the embedding representing the electrons in the Graph Neural Network (GNN) describing the molecular system. 
In this paper, we implement this appending a one-hot encoding of $n$ to the electronic embeddings of the PauliNet2 ansatz~\cite{Schaetzle2023Deep}.

\begin{figure*}[!t]
\begin{center}
\includegraphics[width=15cm]{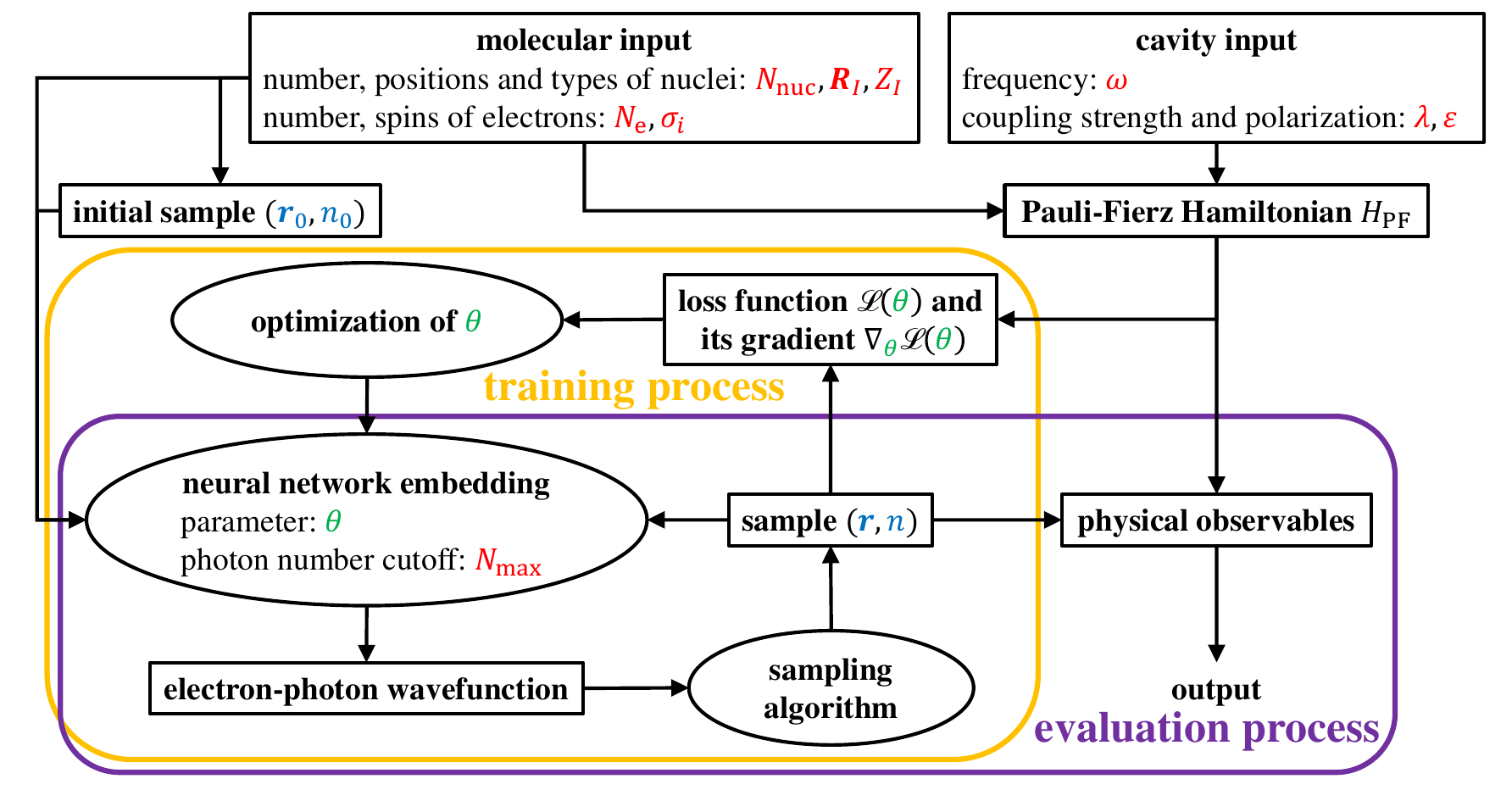}
\end{center}
\caption{\textbf{Flowchart of the variational QMC method.} The input parameters are colored red, the samples of electron positions $\mathbf{r}$ and photon number $n$ are in blue, and the trainable parameters of the neural network are in green. First, an initial batch of samples of $\mathbf{r}$ and $n$ is generated using an educated guess based on the molecular input parameters. The extended neural network produces the corresponding value of the wavefunction. The sampling process uses the Metropolis-Hastings algorithm presented in Sec.~\ref{subsec:dc_algo} to generate more samples of $(\mathbf{r},n)$ distributed according to the squared wavefunction. In the training process (orange region), we optimize the neural network parameter according to the loss function $\mathcal{L}(\theta)$ and its gradient $\nabla_\theta\mathcal{L}(\theta)$.
In the evaluation process (purple region), we use Monte Carlo to estimate the physical observables and output the results.}
\label{fig:flowchart}
\medskip
\small
\end{figure*}

\subsection{Monte Carlo evaluation of physical observables and of the loss function}\label{subsec:obs_el}
In this section, we introduce a Monte Carlo method to evaluate physical observables, the loss function $\mathcal{L}(\theta)$ and its gradient $\partial_\theta\mathcal{L}(\theta)$ that is necessary to perform the gradient-based minimization of the loss function in Eq.~\eqref{eq:loss_function_ext}. 
We consider a general physical observable represented by a Hermitian operator $\mathcal{O}$, which is real and diagonal in its representation under the positional basis $|\mathbf{r}\rangle$, i.e.,
\begin{equation}\label{eq:diagonal_operator_ph}
\langle\mathbf{r},n|\mathcal{O}|\mathbf{r}',m\rangle={O}(\mathbf{r},n,m)\delta(\mathbf{r}-\mathbf{r}'),
\end{equation}
\noindent for some function ${O}(\mathbf{r},n,m)={O}(\mathbf{r},m,n)$. In general, we are interested in evaluating matrix elements of the observable:
\begin{equation}\label{eq:expectation_observable_def}
\begin{aligned}
\langle\mathcal{O}\rangle_{k,l}=&\frac{\langle\psi^{(k)}|\mathcal{O}|\psi^{(l)}\rangle}{\sqrt{\langle\psi^{(k)}|\psi^{(k)}\rangle}\sqrt{\langle\psi^{(l)}|\psi^{(l)}\rangle}},
\end{aligned}
\end{equation}
\noindent where $|\psi^{(k)}\rangle$ with $k\in\{0,1,\cdots,N_\text{s}\}$ is the wavefunction approximating the $k$-th lowest state. When $k=l$, the quantity $\langle\mathcal{O}\rangle_{k,k}=\frac{\langle\psi^{(k)}|\mathcal{O}|\psi^{(k)}\rangle}{\langle\psi^{(k)}|\psi^{(k)}\rangle}$ is the normalized expectation value of $\mathcal{O}$ in $|\psi^{(k)}\rangle$. As shown in App.~\ref{app:obs_ph}, we can evaluate Eq.~\eqref{eq:expectation_observable_def} by using
\begin{eqnarray}
\label{eq:expectation_observable_ph}
\langle\mathcal{O}\rangle_{k,l}
&=&\text{sgn}\left(\mathbb{E}_k\left[{O}_{l,k}(\mathbf{r},n)\right]\right)\nonumber\\
&&\times\sqrt{\mathbb{E}_l\left[{O}_{k,l}(\mathbf{r},n)\right] 
\mathbb{E}_k\left[{O}_{l,k}(\mathbf{r},n)\right]},
\end{eqnarray}
\noindent where 
\begin{eqnarray}
\label{eq:ek}
\mathbb{E}_k[\cdot]&=&\underset{(\mathbf{r},n)\sim|\psi^{(k)}_\theta(\mathbf{r},n)|^2}{\mathbb{E}}[\cdot]\nonumber
\\&=&\sum\limits_{n=0}^{N_{\text{max}}}\int\mathrm{d}\mathbf{r}\frac{|\psi^{(k)}_\theta(\mathbf{r},n)|^2}{\sum\limits_{m=0}^{N_{\text{max}}}\int\mathrm{d}\mathbf{r}''|\psi^{(k)}_\theta(\mathbf{r}'',m)|^2}[\cdot]
\end{eqnarray}
\noindent represents an expectation value over the continuous variable $\mathbf{r}$ and the discrete value $n$ sampled from $\sim |\psi^{(k)}(\mathbf{r},n)|^2$, and
\begin{equation}
\label{eq:olk}
{O}_{l,k}(\mathbf{r},n)=\frac{\sum\limits_{m=0}^{N_{\text{max}}}{O}(\mathbf{r},n,m)\psi^{(l)}(\mathbf{r},m)}{\psi^{(k)}(\mathbf{r},n)}.
\end{equation}
Notice that these expressions generalize the standard ones used in QMC for electronic calculations. Indeed, the standard expectation value over electronic positions $\mathbf{r}$ sampled from $|\psi(\mathbf{r})|^2$~\cite{Hermann2020Deep, Pfau2020Ab, vonGlehn2023A} are now replaced in Eq.~\eqref{eq:ek} with an expectation over $\mathbf{r}$ and $n$ jointly sampled from $|\psi(\mathbf{r},n)|^2$. Furthermore, choosing the electronic Hamiltonian as observable $\mathcal{O}$, we see that Eq.~\eqref{eq:olk} generalizes the standard expression of the ``local energy'' used in QMC~\cite{Hermann2020Deep, Pfau2020Ab, vonGlehn2023A} by introducing an additional sum over the photon number.

Choosing $\mathcal{O} = H_{\text{PF}}$, $S^2$ and $\mathbb{I}$, we can compute respectively the energy of the states, the squared magnitude of the spin and the overlap between different states, which allows us to compute the loss function in Eq.~\eqref{eq:loss_function_ext}. The expression of the loss function, its gradient, and other relevant physical observables are detailed in App.~\ref{app:obs_ph}. 

\subsection{Discrete-continuous Metropolis-Hastings algorithm}\label{subsec:dc_algo}
To efficiently evaluate the loss function and physical observables using Monte Carlo, we replace the expectation values $\mathbb{E}_k[\cdot]$ with an average over finite samples of $(\mathbf{r}, n$) drawn from the probability distribution $\sim|\psi^{(k)}(\mathbf{r},n)|^2$. 
Notice that the state space contains both a continuous and a discrete variable, thus we propose a discrete-continuous version of the traditional Metropolis-Hastings algorithm~\cite{Metropolis1953Equation,Hastings1970Monte}. 
As detailed in App.~\ref{app:sampling}, we choose a proposal function $g((\mathbf{r}',n')\mid(\mathbf{r},n))$ of both variables $(\mathbf{r},n)$ that either updates $\mathbf{r}$ or $n$ with a $50\%$ probability each. If the continuous variable is updated, we propose its change using a continuous proposal function $g_r(\mathbf{r}'\mid \mathbf{r})$ given by a tunable normal distribution, otherwise we use a tunable discrete proposal function $g_n(n'\mid n)$.
The width of these distributions is then adaptively updated to achieve an optimal acceptance ratio \cite{Gelman1997Weak,Roberts2001Optimal}.
Mathematically, this is equivalent to have the proposal function 
\begin{equation}\label{eq:transition_p}
g((\mathbf{r}',n')\mid(\mathbf{r},n))=\frac12\delta_{n,n'}g_r(\mathbf{r}'\mid \mathbf{r})
+\frac12\delta\left(\mathbf{r}-\mathbf{r}'\right)g_n(n'\mid n).
\end{equation}
Once the new values of $(\mathbf{r}', n')$ are proposed, we use the standard Metropolis-Hastings algorithm to perform the update based on the acceptance ratio. The exact algorithm is detailed as Alg.~\ref{alg:2-stepMH} in App.~\ref{app:sampling}.

\subsection{Overview of the method}\label{subsec:summary}
As an overview, the flowchart of our method is shown in Fig.~\ref{fig:flowchart}. The input parameters are denoted in red, including the molecular setting, the cavity parameters, and an ansatz-related photon number cutoff. These parameters are used to calculate the Pauli-Fierz Hamiltonian and to generate an initial batch of samples of electronic positions and photon numbers. The DNN is used to evaluate the wavefunction, and the discrete-continuous sampling algorithm generates new samples (colored blue) to estimate the loss function and its gradient by Monte Carlo evaluation. During the training process (orange region), we use the Kronecker-factored approximate curvature (K-FAC) method~\cite{Martens2015Optimizing} to optimize the loss function and update the neural network parameter $\theta$ (colored green) describing the ansatz. After a sufficient number of training steps, we switch to the evaluation process (purple region) and use Monte Carlo evaluation to estimate and then output the relevant physical observables.

\section{Results}\label{sec:results}
We now demonstrate the capabilities of our deep QMC approach studying one and two hydrogen molecules trapped in an optical cavity. In both cases, we validate our approach showing that our predicted energies are consistent with previous calculations reported in the literature using a coupled-cluster method~\cite{Haugland2020Coupled,Haugland2021Intermolecular} and provide intuitive explanations for the observed features. We then compute additional system observables highlighting the effect of the electron-photon coupling.

First we investigate how the cavity parameters influence the ground-state properties of a two-hydrogen system, focusing on energy, electron density, dipole fluctuations and the photonic state by computing its Wigner function. We observe a change of the energy depending on the polarization of the cavity, a shift in the electron density and decrease of dipole fluctuations compared to the case without a cavity, and a squeezing effect on the electric field. We then study both the ground and excited states properties of a hydrogen molecule and the corresponding photonic states.
As expected, when the cavity frequency matches the energy gap between the ground and the first excited state (in resonance), the effect of the cavity is the strongest and leads to polaritons, i.e. hybrid photonic and electronic states that are a superposition of different photonic and electronic states. Finally, we compute the entanglement entropy between the electronic and photonic degrees of freedom, finding that it is maximal in the resonant case.

\subsection{Ground state properties of two H\textsubscript{2} molecules in a cavity}
\begin{figure}[h]
\centering
\includegraphics[width=8.6cm]{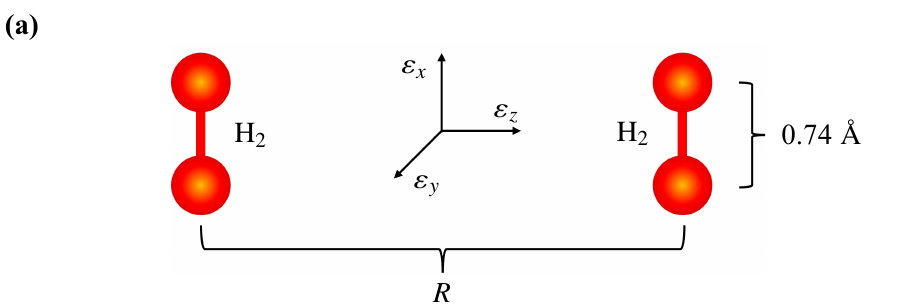}
\includegraphics[width=8.6cm]{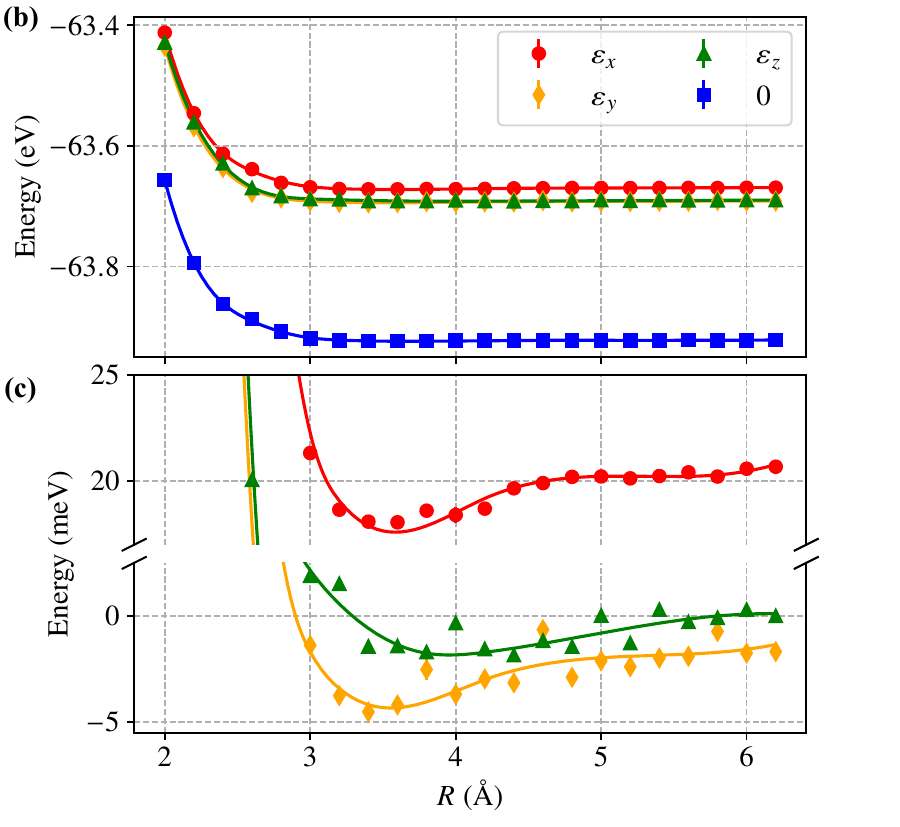}
\caption{\textbf{Physical setting and ground state energies of a two-H\textsubscript{2} system.} \textbf{(a)} Two hydrogen molecules are aligned parallel to each other and separated by distance $R$. The three spatial directions are: $x$ along the molecular axis, $z$ along the separation between two molecules, and $y$ the remaining direction. \textbf{(b)} Ground state energy of two hydrogen molecules as a function of $R$. Each color corresponds to a different polarization of the cavity, as reported in the legend, and the blue curve corresponds to the case without a cavity. The solid curves are fitted using a cubic B-spline. \textbf{(c)} Ground state energy difference between different polarization directions, setting the energy of polarization $\boldsymbol{\varepsilon}_z$ at $R=6.2$ {\AA} as an offset. The energy gap 
between the $\boldsymbol{\varepsilon}_x$ and the $\boldsymbol{\varepsilon}_z$ case is comparable with the results calculated by QED-FCI and QED-CCSD-1~\cite{Haugland2021Intermolecular}, and the sampling error is smaller than the symbol size. Each marked data point is the minimum among five distinct independent runs and the hyperparameters are presented in App.~\ref{app:hyperparameters}.}
\label{fig:H22_energy}
\medskip
\small
\end{figure}

We analyze the molecular system depicted in Fig.~\ref{fig:H22_energy}(a). Two H\textsubscript{2} molecules, each aligned along the $x$-axis with a bond distance of 0.74 {\AA}, are separated along the $z$ by a variable distance $R$.
The origin is chosen as the geometric center of the system. To allow comparison to the literature, we chose the same cavity parameters as Ref.~\cite{Haugland2021Intermolecular}, i.e. a cavity frequency of $\omega = 12.7 \text{ eV}$, and coupling strength $\lambda = 0.1$. We then consider four scenarios, corresponding to the transverse polarization vector $\boldsymbol{\varepsilon}$ of the cavity being along the $x$, $y$ or $z$ axis, and the case without a cavity. 
The direction of the polarization vector depends on the physical implementation of the cavity. For instance, in cavities known as split-ring resonators \cite{maissen2014ultrastrong}, the direction of the polarization has a geometrical meaning, corresponding to the axis where the electric field is confined by the metallic mirrors.

\subsubsection{Ground state energy}
First, we validate our method by calculating the ground state energy of the system and compare it to the results presented in Ref.~\cite{Haugland2021Intermolecular}. In Fig.~\ref{fig:H22_energy}(b), we plot the ground state energies of the two-H\textsubscript{2} molecules as a function of $R$, and each curve corresponds to the four cases mentioned above (the blue curve with squares corresponding to the case without a cavity). We perform a separate optimization for each value of $R$ between $2$ {\AA} and $6.2$ {\AA} with a 0.2 {\AA} spacing.

As we can see from Fig.~\ref{fig:H22_energy}(b), on this scale the ground state energy decreases as $R$ increases in all four cases. In the presence of the cavity, we observe an energy increase of more than 200 meV. The increase is larger when the polarization is parallel with the hydrogen bond ($\boldsymbol{\varepsilon}_x$ case) than when it is in the direction of the separation of two molecules ($\boldsymbol{\varepsilon}_z$ case). This is due to larger dipole fluctuations along the $x$ direction, leading to a larger dipole self-energy term in Eq.~\eqref{eq:Hamiltonian_all}.
In Fig.~\ref{fig:H22_energy}(c) we take a closer look at the energy difference between different cavity polarizations. The energies in the $\boldsymbol{\varepsilon}_y$ case (polarization along the $y$-axis) are slightly lower than those of $\boldsymbol{\varepsilon}_z$, while the energy in the $\boldsymbol{\varepsilon}_x$ case is larger than in the $\boldsymbol{\varepsilon}_z$ case by about $22$ meV within the shown range. This energy gap signals that the hydrogen molecules in a cavity tend to align themselves such that the cavity polarization is not along the $x$ direction, i.e. the molecular bond direction.

To validate our method, we compare the results with the calculations performed in Ref.~\cite{Haugland2021Intermolecular}, which studies the same system using two different methods: QED-CCSD-1 and QED-FCI. With both methods above, they also find that the energies of $\boldsymbol{\varepsilon}_y$ are lower than those in the $\boldsymbol{\varepsilon}_z$ case. The energy gap between $\boldsymbol{\varepsilon}_x$ and $\boldsymbol{\varepsilon}_z$ at $R=6.0$ {\AA} is around $40$ meV using QED-CCSD-1 and is around $34$ meV using QED-FCI, which is expected to be more accurate than QED-CCSD-1. These results are comparable to the value of the gap that we find ($22$ meV), confirming the validity of our approach. 
We also reproduce the energy minimum at $R\approx3.5$ {\AA} in all three cases. It is worth mentioning that the height of the ``dip'' is about 3 meV, which is 0.07 kcal/mol and is much smaller than the error bar of the results from some cutting-edge deep QMC methods~\cite{Entwistle2023Electronic,vonGlehn2023A}.

\subsubsection{Electron density of the ground state}\label{sec:rho_e}
\begin{figure}[t]
\centering
\includegraphics[width=8.6cm]{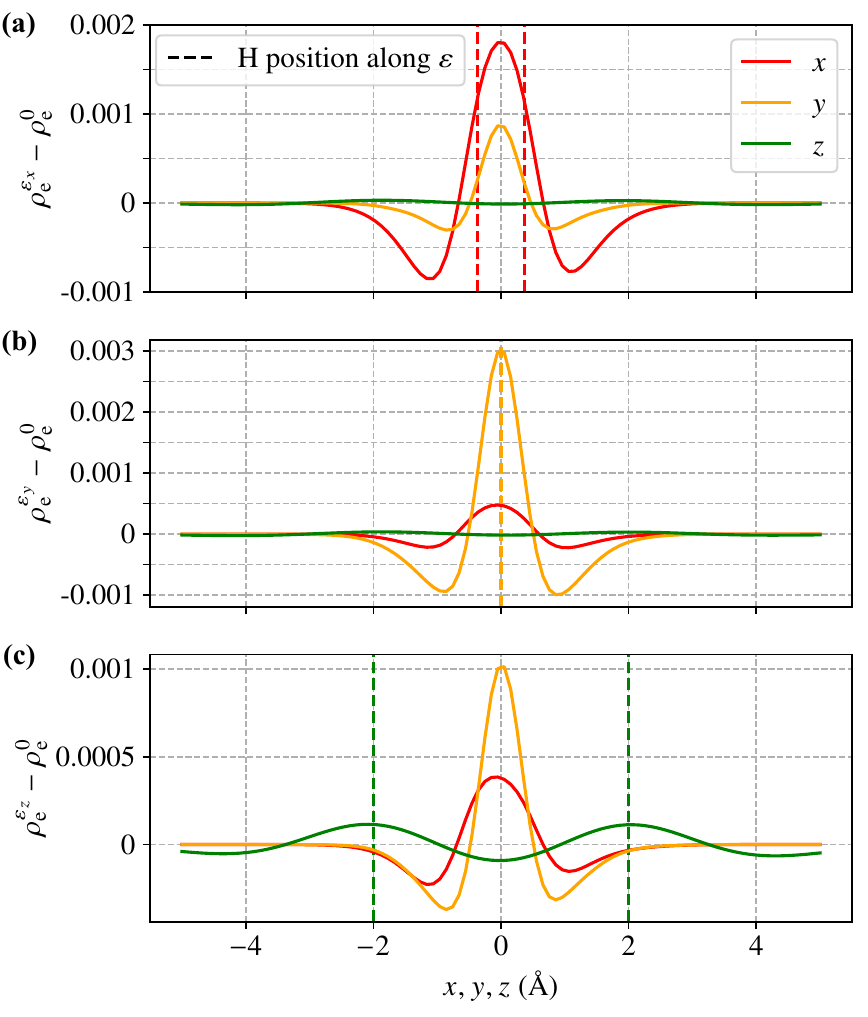}
\includegraphics[width=8.6cm]{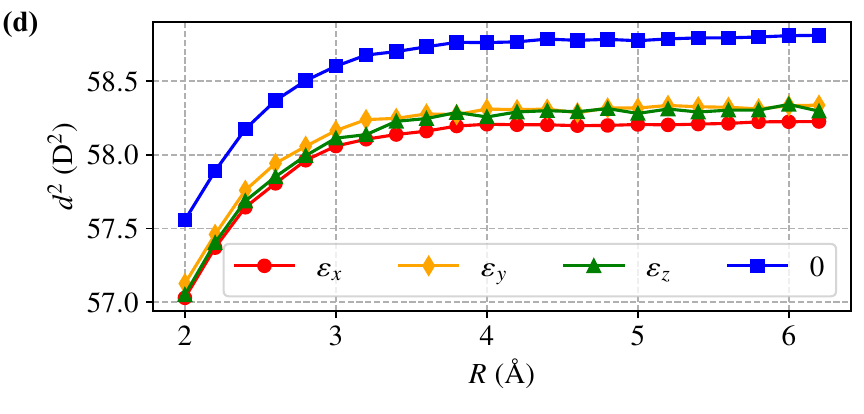}
\caption{\textbf{Cavity effect on electron densities and dipole flucutations.} The data points are calculated from the same states as in Fig.~\ref{fig:H22_energy}.
\textbf{(a)} -- \textbf{(c)} Electron density difference between the molecular system with $R=4.0$ {\AA} in a cavity with different polarizations ($\rho_\text{e}^{\boldsymbol{\varepsilon}_x}$, $\rho_\text{e}^{\boldsymbol{\varepsilon}_y}$ and $\rho_\text{e}^{\boldsymbol{\varepsilon}_z}$) and that in the absence of a cavity ($\rho_\text{e}^{0}$), observed along three spatial dimensions reported in the legend. Colored dashed lines are the coordinates of the two nuclei in the direction of the polarization. The main effect of the cavity is to concentrate electrons around the nuclei of each molecule. \textbf{(d)} Fluctuations of electronic dipole moment (in Debye units) in the ground state as a function of $R$ for each polarization case and in the absence of the cavity. Fluctuations are lowered in the cavity compared to the case without a cavity.}
\label{fig:H22_shift_and_dipole}
\medskip
\small
\end{figure}

Our approach provides us with the full electron-photon wavefunction, allowing us to calculate any system observable. To visualize how the optical cavity with different polarizations affects the system, we analyze the electron density $\rho_\text{e}(\vec{r})$, i.e. the (normalized) probability of an electron being present at a spatial point $\vec{r}=(x,y,z)$. We estimate this quantity according to the formula
\begin{equation}\label{eq:electron_density}
\begin{aligned}
\rho_\text{e}(\vec{r})&=\mathbb{E}_{(\mathbf{r},n)\sim|\psi(\mathbf{r},n)|^2}\left[\frac1{N_{\text{e}}}\sum\limits_{i=1}^{N_\text{e}}\delta(\vec{r}-\mathbf{r}_i)\right]
\end{aligned}
\end{equation}
\noindent using Monte Carlo and then using Gaussian kernel density estimation~\cite{Scott2015Multivariate} (see App.~\ref{app:electron_density} for details).
In Fig.~\ref{fig:H22_shift_and_dipole} we compute the charge displacement, i.e. the difference between the electron density $\rho_\text{e}^{\epsilon_i}(\vec{r})$ in a cavity with polarization $\epsilon_i$, for $i=x,y,z$, and the density $\rho_\text{e}^0(\vec{r})$ in the absence of the cavity. Each panel (a), (b), and (c) corresponds to a different cavity polarization setting $R=4.0$ {\AA}. 
Each curve, instead, corresponds to the axis (reported in the legend) along which we observe the charge displacement; mathematically, this corresponds to integrating the electron density over the other two dimensions. For example, $\rho_\text{e}(x) = \int\int \rho_\text{e}(\vec{r}) \dd y \dd z$. 

The main effect of the cavity is to concentrate the electronic density around the center of each molecule. Indeed, in the $\boldsymbol{\varepsilon}_x$ case, reported in Fig.~\ref{fig:H22_shift_and_dipole}(a), we observe an increase of the electron density between the nuclei (peak of the red curve in the origin), which are located at $x=\pm 0.37$ \text{\AA} (dashed red lines). Correspondingly, there is a charge depletion further away from the nuclei. The concentration of the charge around the nuclei is also visible in the $y$ and $z$ directions (orange and green curves), albeit less pronounced.
In the $\boldsymbol{\varepsilon}_y$ case, reported in Fig.~\ref{fig:H22_shift_and_dipole}(b), the charge displacement occurs almost exclusively along the $y$ axis, where the electronic density is concentrated in the origin, i.e. over the nuclei (dashed yellow line).
In the $\boldsymbol{\varepsilon}_z$ case, reported in  Fig.~\ref{fig:H22_shift_and_dipole}(c), the electrons concentrate around $z=\pm2$ {\AA}, corresponding to the nuclei position (green dashed lines), and are depleted from the surrounding areas. The gathering effect around the nuclear position is also visible along the $y$ axis, but it is less pronounced than in the $\boldsymbol{\varepsilon}_y$ case.

These results are consistent with our intuition since the electron-photon interaction couples the photons to the molecule's dipole along the polarizing direction, and thus the largest effect is expected to be indeed along this axis. To further understand this effect, in Fig.~\ref{fig:H22_shift_and_dipole}(d) we plot the fluctuations of electronic dipole moment $d^2$ as a function of $R$ in the three polarization cases, and in the absence of the cavity (notice that the average dipole moment is zero due to the symmetry of the system).
Interestingly, we notice that the fluctuations are lower in the cavity compared to the case without a cavity, with the reduction being maximum in the $\boldsymbol{\varepsilon}_x$ case. Indeed,  fluctuations of the dipole create a fluctuation in the electric field, which in turn requires energy to create. Hence, to minimize the overall energy for the ground state, the cavity tends to decrease the fluctuations of the molecules' dipole.

\subsubsection{Photonic state}\label{sec:photon_wf}
We now analyze the state of the photonic system. Fixing the distance between two molecules at $R=4.0\text{ \AA}$, we calculate the photonic reduced density matrix 
\begin{equation}
    \rho_{\text{ph}}^{\boldsymbol{\varepsilon}_x} = \frac{\mathrm{Tr}_{\mathbf{r}}(|\psi\rangle\langle\psi|)}{\mathrm{Tr}(|\psi\rangle\langle\psi|)}
\end{equation} 
of the ground state in the cavity with polarization $\boldsymbol{\varepsilon}_x$
using the expressions in App.~\ref{app:photon_dm}, $\mathrm{Tr}_{\mathbf{r}}(\cdot)$ denoting the partial trace over the electronic coordinates, and $\mathrm{Tr}(\cdot)$ the full trace. The full numerical density matrix is reported in App.~\ref{app:wigner}.
Analyzing the diagonal part of the matrix expressed in the Fock basis, we find that the probability of finding $n=0$ photons is very large ($>0.99$), and the following terms (representing the probability of finding $n>0$ photons)
decrease rapidly ($<0.008$ for $n=1$ and $<2\times10^{-4}$ for $n=2$), justifying our assumption of a maximum number of photons in the wavefunction ansatz.
The off-diagonal terms, which reach at most the value of 0.01, represent coherence among Fock states with different photon numbers. As a reference, in the absence of a cavity, the cavity is in a pure state with $n=0$ photons, therefore with no coherence between Fock states.

To get a better understanding of the density matrix of the photonic state, we calculate the Wigner function~\cite{Wigner1932On} of the photonic state $\rho_{\text{ph}}^{\boldsymbol{\varepsilon}_x}$ and compare it to that of the photonic ground state $|0\rangle$. 
For a quantum state with density matrix $\rho$, it is defined as
\begin{equation}\label{eq:wigner}
W(x,p)=\frac1{\pi}\int_{-\infty}^{+\infty}\langle x+y|\rho|x-y\rangle e^{-2ipy}\mathrm{d}y,
\end{equation}
where $\ket{x}$ (resp. $\ket{p}$) is the set of eigenstates of the operator $\hat{x}$ (resp. $\hat{p})$, defined as
\begin{eqnarray}\label{eq:wigner_xp}
\hat{x}&=&\frac1{\sqrt{2}}(\hat{b}+\hat{b}^\dagger),\\
\hat{p}&=&\frac i{\sqrt{2}}(-\hat{b}+\hat{b}^\dagger).
\end{eqnarray}
In the dipole gauge, the operators $\hat{x}$ and $\hat{p}$ are respectively proportional to the transverse displacement field $\mathbf{D}_\perp$ and the vector potential $\mathbf{A}$ projected on the polarization vector $\boldsymbol{\varepsilon}$ \cite{Schaefer2020Relevance}.

\begin{figure}[t]
\centering
\includegraphics[width=8.6cm]{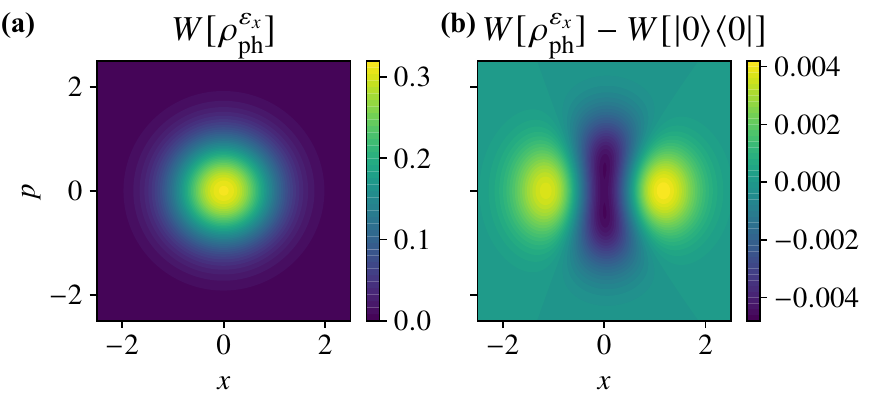}
\caption{\textbf{Photonic system of the ground state.} \textbf{(a)} Wigner function of the photonic state $\rho_{\text{ph}}^{\boldsymbol{\varepsilon}_x}$, at with $R=4.0$ {\AA}, as a function of $x$ and $p$ which are, respectively, proportional to the transverse displacement field $\mathbf{D}_\perp$ and to the vector potential $\mathbf{A}$. It is almost identical to the Wigner function of $|0\rangle$. \textbf{(b)} The difference between Wigner function of $\rho_{\text{ph}}^{\boldsymbol{\varepsilon}_x}$ and that of the Fock state $|0\rangle\langle0|$. 
The main effect of the electron-photon coupling is to increase the fluctuations of $\mathbf{D}_\perp$ (proportional to $x$).}
\label{fig:wigner_x}
\medskip
\small
\end{figure}

Fig.~\ref{fig:wigner_x}(a) shows the Wigner function of the photonic state $\rho_{\text{ph}}^{\boldsymbol{\varepsilon}_x}$. Since the state $\rho_{\text{ph}}^{\boldsymbol{\varepsilon}_x}$ is very close to the Fock state $|0\rangle$ (corresponding to the no-cavity case), its Wigner function is almost identical to a two-dimensional Gaussian distribution~\cite{Serafini2017Quantum}.
To visualize what is the effect of the cavity on the Wigner function, we then calculate the difference between $W[\rho_{\text{ph}}^{\boldsymbol{\varepsilon}_x}]$ and $W[|0\rangle\langle0|]$ and plot it in Fig.~\ref{fig:wigner_x}(b). We observe that the Wigner function is squeezed along $p$, leading to an increase along $x$. Since the transverse displacement field $\mathbf{D}_\perp$ is proportional to $x$, the electron-photon interaction has the effect of increasing the fluctuations of $\mathbf{D}_\perp$. Indeed, the probability of $\mathbf{D}_\perp$ being zero decreases (blue region), and that of taking values corresponding to $|x|\in[1,1.5]$ increases (yellow regions). We could find a qualitative explanation of this squeezing effect using a toy model and a second-order perturbative expansion in the electron-photon coupling, see App.~\ref{app:squeezing} for details. 

\subsection{Ground and excited states of H\textsubscript{2}}
We now study the properties of the ground and excited states of an H\textsubscript{2} molecule inside a cavity [see Fig~\ref{fig:H2_energy}(a)]. As in Ref.~\cite{Haugland2020Coupled}, we choose the polarization to be parallel to the direction of two nuclei with strength $\lambda=0.05$. The frequency is set to be $\omega=12.75$ eV so that it is in resonance with the first bright excitation of H\textsubscript{2} at the equilibrium geometry, where the two nuclei are separated by $R_0=0.74$ {\AA}~\cite{Kolos1965Potential}.

\subsubsection{Excited state energies}
\begin{figure}[t]
\centering
\includegraphics[width=8.6cm]{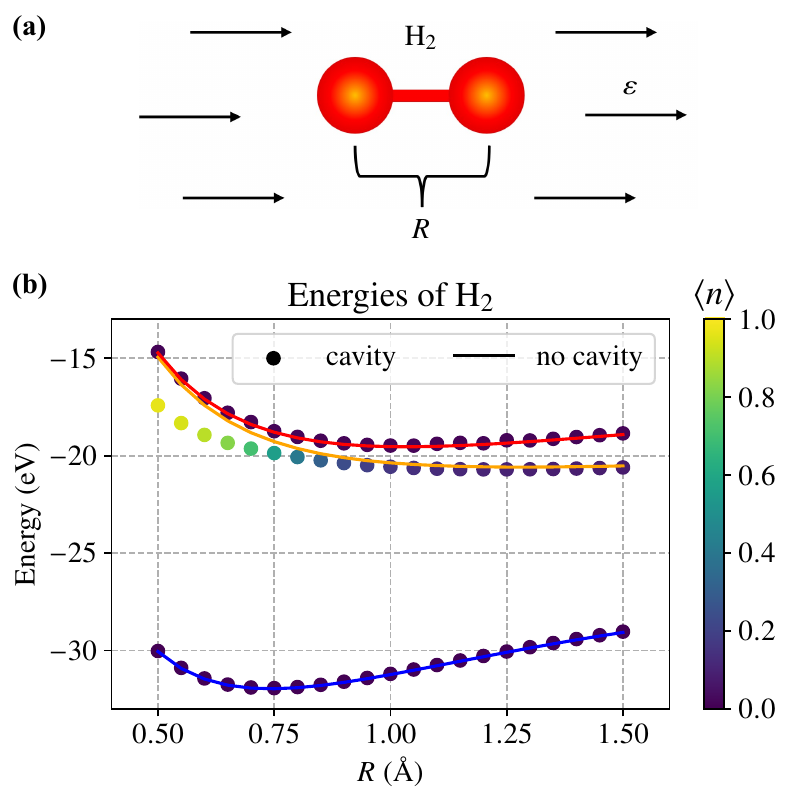}
\caption{\textbf{Physical setting, energies, and average photon numbers of a H\textsubscript{2} system.} \textbf{(a)} Two hydrogen atoms are aligned parallel to the cavity polarization and separated by distance $R$. As in Ref.~\cite{Haugland2020Coupled}, the cavity frequency is set to be in resonance with the first bright excitation at its equilibrium geometry, and the coupling strength is $\lambda=0.05$. \textbf{(b)} Ground state and excited state energies of a hydrogen molecule as a function of the distance $R$ between the nuclei with (dots) and without (lines) the cavity. Only states with zero spin (singlets) are reported. Each dot's color indicates the average photon number of that state. Each data point is obtained by using the hyperparameters in App.~\ref{app:hyperparameters}. A full optimization at all distances was repeated 6 times, and one of the two runs that did not present any outlier energies is reported in the figure.}
\label{fig:H2_energy}
\medskip
\small
\end{figure}

In Fig.~\ref{fig:H2_energy}(b), we plot the energy of the first three singlet states as a function of the distance $R$ between the two nuclei, both with (dots) and without (lines) the cavity. The average photon number $\ev*{n}$ is shown as the color of each dot.
As we can see, the ground state energies are not affected by the cavity, so the ground state of the system is approximately the tensor product of the electronic ground state $|\psi_\text{e}^{(0)}\rangle$ (blue curve) and photonic ground state $|0\rangle$, 
\begin{equation}
|\psi^{(0)}\rangle\approx |\psi_\text{e}^{(0)}\rangle\otimes|0\rangle,
\end{equation} 
where coefficients of higher excitations are very small. 

Conversely, we observe a strong impact of the cavity on the first excited state. 
To qualitatively understand this behavior, let us expand the first excited state in the cavity as a superposition of states with either a single excitation in the electronic system $|\psi_\text{e}^{(1)}\rangle\otimes|0\rangle$, or a single excitation in the photonic system $|\psi_\text{e}^{(0)}\rangle\otimes|1\rangle$, i.e.
\begin{equation}
\label{eq:first_excited_superposition}
|\psi^{(1)}\rangle\approx c_0|\psi_\text{e}^{(1)}\rangle\otimes|0\rangle + c_1|\psi_\text{e}^{(0)}\rangle\otimes|1\rangle.
\end{equation}
Since the cavity is resonant with the first excitation at the equilibrium geometry $R_0=0.74$ {\AA}, inducing an excitation in the electronic or photonic system at $R=R_0$ provides the same energy, so we expect $|c_0|^2 = |c_1|^2=1/2$. Indeed, as we can see in Fig.~\ref{fig:H2_energy}, the average photon number at $R$ close to $R_0$ is $\ev{n}\approx1/2$. As we move to smaller values of $R$, the energy of $|\psi_\text{e}^{(0)}\rangle\otimes|1\rangle$ becomes smaller than $|\psi_\text{e}^{(1)}\rangle\otimes|0\rangle$, so $|c_1| > |c_0|$. Therefore, the average photon number increases towards $1$, and the electronic state becomes closer to the electronic ground state $\ket*{\psi_\text{e}^{(0)}}$. Conversely, as $R$ increases, the cost of exciting the electronic system is lower than that of creating a photon, so $\ev{n}$ decreases towards zero, and the electronic state tends to the first electronic excited state $\ket*{\psi_\text{e}^{(1)}}$; indeed, the dots associated with the first excited state overlap with the orange curve for large $R$. The second excited state, instead, only has black dots, and it overlaps with the red curve, so its state is approximately $|\psi_\text{e}^{(2)}\rangle\otimes|0\rangle$.

We compare Fig.~\ref{fig:H2_energy}(b) to the potential energy curves of H\textsubscript{2} in Ref.~\cite{Haugland2020Coupled}, where the same system is studied using QED-CCSD-1.
We find that the energies of $|\psi^{(0)}\rangle$ and $|\psi^{(1)}\rangle$ are the same for both methods, confirming the validity of our approach. 
However, the results for $|\psi^{(2)}\rangle$ differ: while in Ref.~\cite{Haugland2020Coupled} it is a superposition of $|\psi_\text{e}^{(0)}\rangle\otimes|1\rangle$ and $|\psi_\text{e}^{(1)}\rangle\otimes|0\rangle$, we find here that $|\psi^{(2)}\rangle$ is approximately $|\psi_\text{e}^{(2)}\rangle\otimes|0\rangle$. This discrepancy arises from the mismatch between the energy of $|\psi_\text{e}^{(2)}\rangle$ computed with QMC and CCSD. Specifically, in Ref.~\cite{Haugland2020Coupled}, the energy gap between $|\psi_\text{e}^{(2)}\rangle$ and $|\psi_\text{e}^{(1)}\rangle$ is several eVs, while it is much smaller in our description.
Consequently, when the cavity is introduced, the superposition state predicted by Ref.~\cite{Haugland2020Coupled} has a lower energy than the state $|\psi_\text{e}^{(2)}\rangle\otimes|0\rangle$, while this energy ordering is reversed in our case.
Our second excited state thus corresponds to approximately $|\psi_\text{e}^{(2)}\rangle\otimes|0\rangle$, and not to the superposition state found in Ref.~\cite{Haugland2020Coupled}.

To address the energy discrepancy of $|\psi_\text{e}^{(2)}\rangle$ between our work and Ref.~\cite{Haugland2020Coupled}, we analyze the energy curves of H\textsubscript{2} using the CASSCF method with varying basis-set sizes (see App.~\ref{app:CASSCF_H2}). Using a small basis-set (cc-pVDZ, as used in Ref.~\cite{Haugland2020Coupled}), we recover their description. However, as we increase the size of the basis-set (and so the accuracy), we converge to the QMC results, confirming the validity of our approach.

\subsubsection{Photonic state}\label{sec:photon_wf_e}
\begin{figure}[t]
\centering
\includegraphics[width=8.6cm]{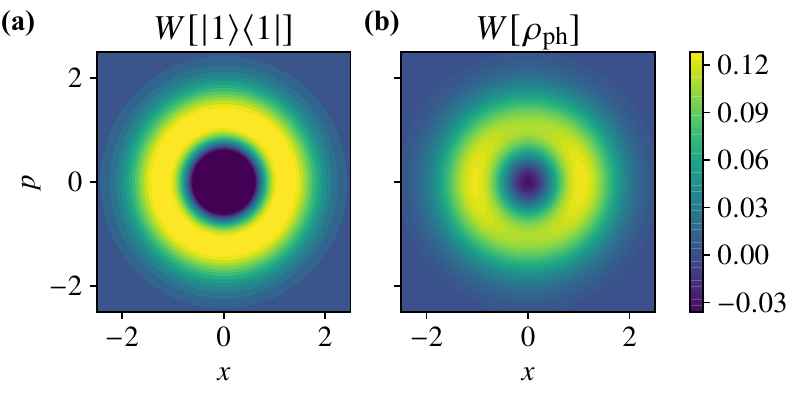}
\caption{\textbf{Photonic system of the first singlet excited state.} \textbf{(a)} Wigner function of $|1\rangle$. \textbf{(b)} Wigner function of the photonic state $\rho_{\text{ph}}$, reported in Eq.~\eqref{eq:photon_dm_e}, corresponding to the first excited state at $R=0.75$ {\AA}. The state is close to a superposition of $|0\rangle$ and $|1\rangle$, with comparable coefficients. The values in the area around its center are negative, which is a characteristic of $W[|1\rangle\langle1|]$.}
\label{fig:wigner_excited}
\medskip
\small
\end{figure}

As discussed in the previous section, the first singlet excited state $|\psi^{(1)}\rangle$ is a hybrid light-matter state that transitions from a more ``photonic'' to a more ``electronic'' character as $R$ increases. We pick $R=0.75$ {\AA}, which is close to its equilibrium geometry $R_0=0.74$ {\AA} without a cavity, and estimate the density matrix of the photonic state in this case (the numerical density matrix is reported in App.~\ref{app:wigner}.)
The fact that the first two diagonal entries are $\simeq 1/2$ confirms that indeed $|c_0|\approx |c_1|$ in Eq.~\eqref{eq:first_excited_superposition}, meaning the state is close to a superposition of the $|0\rangle$ and $|1\rangle$ photonic states. The corresponding Wigner function is shown in Fig.~\ref{fig:wigner_excited}(b). As opposed to the Wigner function of $|0\rangle$, which is very similar to the Winger function reported in Fig.~\ref{fig:wigner_x}(a), the Wigner function of $\rho_{\text{ph}}$ in the central area is negative in the center, which is a characteristic of the Wigner function of $|1\rangle$ [Fig.~\ref{fig:wigner_excited}(a)]. 

\begin{figure}[t]
\centering
\includegraphics[width=8.6cm]{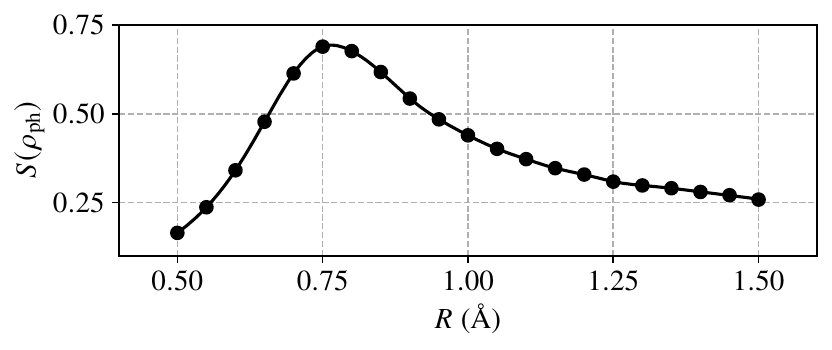}
\caption{\textbf{Entanglement between the electronic and photonic subsystems.} We calculate the von Neumann entropy of $\rho_{\text{ph}}$ as a function of $R$, associated with the first excited state $|\psi^{(1)}\rangle$ [see Eq.~\eqref{eq:entropy}]. It reaches the peak at $R$ close to the equilibrium geometry $R_0$, where the electromagnetic field is in resonance with the first singlet excited state of the molecule.}
\label{fig:entropy}
\medskip
\small
\end{figure}

Finally, in Fig.~\ref{fig:entropy} we report the entanglement between the electronic and photonic subsystems, as a function of $R$, analyzing the first singlet excited state $|\psi^{(1)}\rangle$. To quantify entanglement, we calculate the entanglement entropy~\cite{Horodecki2009Quantum}, which corresponds to von Neumann entropy of the reduced density matrix of one of the two subsystems. 
It is positive only if the two subsystems are entangled.
Here, we choose to compute the entropy of the photonic reduced density matrix, i.e.
\begin{equation}\label{eq:entropy}
S(\rho_{\text{ph}})=-\Tr\left(\rho_{\text{ph}}\ln\rho_{\text{ph}}\right).
\end{equation}
As expected, the entanglement is maximum at the equilibrium geometry, where the electromagnetic field is in resonance with the first singlet excited state $|\psi^{(1)}\rangle$, and decreases as we move away from resonance.

\section{Conclusions}\label{sec:conclusions}
We present an ab-initio approach, based on machine learning and on variational quantum Monte Carlo, to solve the Schr\"odinger equation of molecules strongly coupled to optical cavities. Our approach can compute both the ground and excited states of the light-matter system, allowing us to access any electronic or photonic property of the system. 
In particular, we propose an extension of deep electronic wavefunction ansatzes to describe also the quantum nature of the cavity photons, introducing a Monte Carlo framework that deals with the electrons in first quantization, and with the photons in second quantization.

We then showcase and validate our method using an extension of the PauliNet2 ansatz~\cite{Schaetzle2023Deep} to calculate the ground and excited states of small molecular systems.
First, we compute the ground state energies of two hydrogen molecules in a cavity with different polarization directions. Our method reproduces the results obtained from different quantum chemistry methods \cite{Haugland2021Intermolecular}, with energy differences within 1 kcal/mol. We then analyze the effect of the cavity computing the electronic density displacement caused by the cavity,
and the photonic density matrix and corresponding Wigner function.
We then study a single H\textsubscript{2} molecule in a cavity computing both the ground and excited states, and we compare them with those found using coupled cluster methods~\cite{Haugland2020Coupled}. We find genuine light-matter hybrid excited states that are given by a superposition of different electronic wavefunctions and different photon numbers. We compute the entanglement generated between the electronic and photonic subsystems, finding that it is maximum when the cavity frequency is resonant with the first excitation energy. 

Our method paves the way for the systematic and accurate ab initio study of the effect of optical cavities on small molecular systems, extending the range of scientific problems that can be tackled using machine learning techniques. It also constitutes an example of a neural network wavefunction ansatz that can describe joint fermionic and bosonic systems. As a direction for future research, our method could be extended to a transferable ansatz~\cite{Scherbela2024Towards,Gerard2024Transferable}, i.e. a single neural network ansatz that predicts the wave-function for multiple nuclear positions. This approach could increase the accuracy of dissociation curves through a relative error cancellation between different nuclear configurations. Further strategies to improve the accuracy of the method can be considered, such as the variance matching scheme \cite{Robinson2017Excitation}.

Since our approach gives us access to the full electron-photon wavefunction, we envision using the method to study further chemical properties, such as photochemical reactions under the influence of light-matter interaction~\cite{Thomas2016Ground,Galego2019Cavity,Thomas2019Tilting}. Larger molecules could also be studied, and our approach could help identify systems where the effect of the cavity is most relevant.

Finally, by substituting photons with phonons, a similar ansatz can be used to study the effect of electron-phonon coupling~\cite{Ohgoe2014Variational}.

\section*{Acknowledgements}
We would like to thank Jan Hermann, Mike Entwistle, Marco Polini, Marco Schirò, Enrico Ronca, Angel Rubio, and Filippo Vicentini for helpful discussions and suggestions.
Y. T. acknowledges funding from the European Research Council (ERC) Consolidator Grant ScaleCell. G. M. A. acknowledges funding from the European Union’s Horizon 2020 research and innovation programme under the Marie Sklodowska-Curie (Grant Agreement No. 101146870 --  COMPASS) and from the European Research Council (ERC) under the European Union's Horizon 2020 research and innovation programme (Grant agreement No. 101002955 -- CONQUER). A.C. acknowledges support from the Alexander von Humboldt Foundation.  M. M. acknowledges funding from the European Research Council (ERC) under the European Union’s Horizon 2020 research and innovation program (Grant Agreement No. 772230). P. B. S. and Z. S. acknowledge the Berlin Mathematics
 Research Center MATH+ (Project No. AA2-22 and AA2-8). F. N. gratefully acknowledges funding by the
 BMBF (Berlin Institute for the Foundations of Learning
 and Data—BIFOLD), the European Research Commission
 (ERC CoG 772230) and the Berlin Mathematics Center
 MATH+(AA1-6, AA2-8).
 P. A. E. gratefully acknowledges funding by the Berlin Mathematics Center MATH+ (AA2-18). 

 \section*{Author Declarations}
 \subsection*{Conflict of Interest}
 The authors have no conflicts to disclose.

 \subsection*{Author Contributions}


\textbf{Yifan Tang:} 
Conceptualization (supporting); 
Data curation (lead); 
Formal analysis (equal); 
Methodology (equal); 
Investigation (lead); 
Software (lead); 
Validation (equal); 
Visualization (lead); 
Writing - Original draft preparation (lead); 
Writing - review \& editing (equal)
\textbf{Gian Marcello Andolina:} 
Conceptualization (equal); 
Formal analysis (equal); 
Methodology (equal); 
Validation (equal); 
Project administration (supporting);
Supervision (supporting);
Writing - Original draft preparation (supporting); 
Writing - review \& editing (lead)
Funding acquisition;
\textbf{Alice Cuzzocrea:}
Data curation (supporting);
Formal analysis (equal); 
Visualization (supporting);
Writing - review \& editing (supporting);
Funding acquisition;
\textbf{Mat\v ej Mezera:} 
Methodology (supporting); 
Software (supporting); 
Writing - review \& editing (supporting)
\textbf{P. Bern\'at Szab\'o:} 
Methodology (supporting); 
Software (supporting); 
Writing - review \& editing (supporting)
\textbf{Zeno Sch\"atzle:} 
Methodology (supporting); 
Software (supporting);
 Writing - review \& editing (supporting)
\textbf{Frank No\'e:} 
Conceptualization (equal); 
Methodology (supporting);
Funding acquisition;
Project administration (equal);
Resources;
Supervision (equal);
Writing - review \& editing (supporting);
\textbf{Paolo A. Erdman:}
Conceptualization (equal);
Formal analysis (equal);
Investigation (supporting);
Methodology (equal);
Funding acquisition;
Project administration (equal);
Supervision (equal);
Validation (equal); 
Writing - Original draft preparation (supporting);
Writing - review \& editing (lead);

 \section*{Data Availability Statement}

 The data that support the findings of this study are available from the corresponding author upon reasonable request.

\appendix

\begin{appendices}

\section{Monte Carlo evaluation of physical observables, the loss function and its gradient in electron-photon systems}\label{app:obs_ph}

Here we show how to evaluate observables using Monte Carlo in the electron-photon case, and we present explicit expressions of the loss function, its gradient, and other relevant observables.
We start by proving Eq.~\eqref{eq:expectation_observable_ph} of the main text, which allows to evaluate the expectation value of a general physical observables represented by a Hermitian operator $\mathcal{O}$. 
We assume that it is real and diagonal in its representation under the positional basis $|\mathbf{r}\rangle$, i.e.,
\begin{equation}\label{eq:diagonal_operator_ph}
\langle\mathbf{r},n|\mathcal{O}|\mathbf{r}',m\rangle={O}(\mathbf{r},n,m)\delta(\mathbf{r}-\mathbf{r}'),
\end{equation}
\noindent with ${O}(\mathbf{r},n,m)=\mathcal{O}(\mathbf{r},m,n)$. Furthermore, we assume $\psi(\mathbf{r},n)$ only takes real values. Using Eq.~\eqref{eq:diagonal_operator_ph}, we have
\begin{widetext}
\begin{eqnarray}
\label{eq:obs_kl_ph}
\langle\psi^{(k)}|\mathcal{O}|\psi^{(l)}\rangle 
&=& \sum\limits_{n,m=0}^{N_{\text{max}}} \int\mathrm{d}\mathbf{r}\mathrm{d}\mathbf{r}'\langle\psi^{(k)}|\mathbf{r},n\rangle\langle\mathbf{r},n|\mathcal{O}|\mathbf{r}',m\rangle\langle\mathbf{r}',m|\psi^{(l)}\rangle
 = \sum\limits_{n,m=0}^{N_{\text{max}}}\int\mathrm{d}\mathbf{r}\psi^{(k)}(\mathbf{r},n){O}(\mathbf{r},n,m)\psi^{(l)}(\mathbf{r},m)\nonumber
\\
&=&\langle\psi^{(k)}|\psi^{(k)}\rangle\sum\limits_{n=0}^{N_{\text{max}}}\int\mathrm{d}\mathbf{r}\frac{|\psi^{(k)}(\mathbf{r},n)|^2}{\sum\limits_{m=0}^{N_{\text{max}}}\int\mathrm{d}\mathbf{r}''|\psi^{(k)}(\mathbf{r}'',m)|^2}\frac{\sum\limits_{m=0}^{N_{\text{max}}}{O}(\mathbf{r},n,m)\psi^{(l)}(\mathbf{r},m)}{\psi^{(k)}(\mathbf{r},n)}\nonumber
\\
&=&\langle\psi^{(k)}|\psi^{(k)}\rangle\mathbb{E}_k\left[\frac{\sum\limits_{m=0}^{N_{\text{max}}}{O}(\mathbf{r},n,m)\psi^{(l)}(\mathbf{r},m)}{\psi^{(k)}(\mathbf{r},n)}\right],
\end{eqnarray}
\noindent where $\mathbb{E}_k[\cdot]$ is defined in Eq.~(\ref{eq:ek}) of the main text. Using that $\langle\psi^{(k)}|\mathcal{O}|\psi^{(l)}\rangle=\langle\psi^{(l)}|\mathcal{O}|\psi^{(k)}\rangle$ since $\mathcal{O}$ is real and Hermitian, we have that
\begin{equation}
\langle\mathcal{O}\rangle_{k,l}^2
=\frac{\langle\psi^{(k)}|\mathcal{O}|\psi^{(l)}\rangle}{\langle\psi^{(k)}|\psi^{(k)}\rangle}
\frac{ \langle\psi^{(l)}|\mathcal{O}|\psi^{(k)}\rangle}{\langle\psi^{(l)}|\psi^{(l)}\rangle}.
\end{equation}
\noindent
Plugging in Eq.~(\ref{eq:obs_kl_ph}) yields
\begin{equation}
\langle\mathcal{O}\rangle_{k,l}^2
=\mathbb{E}_k\left[\frac{\sum\limits_{m=0}^{N_{\text{max}}}{O}(\mathbf{r},n,m)\psi^{(l)}(\mathbf{r},m)}{\psi^{(k)}(\mathbf{r},n)}\right]
\mathbb{E}_l\left[\frac{\sum\limits_{m=0}^{N_{\text{max}}}{O}(\mathbf{r},n,m)\psi^{(k)}(\mathbf{r},m)}{\psi^{(l)}(\mathbf{r},n)}\right],
\end{equation}
Taking the square root, it can be shown that
\begin{equation}\label{eq:exp_O_kl}
\langle\mathcal{O}\rangle_{k,l}
=\text{sgn}\left(\mathbb{E}_k\left[\frac{\sum\limits_{m=0}^{N_{\text{max}}}{O}(\mathbf{r},n,m)\psi^{(l)}(\mathbf{r},m)}{\psi^{(k)}(\mathbf{r},n)}\right]\right)
\sqrt{\mathbb{E}_k\left[\frac{\sum\limits_{m=0}^{N_{\text{max}}}{O}(\mathbf{r},n,m)\psi^{(l)}(\mathbf{r},m)}{\psi^{(k)}(\mathbf{r},n)}\right]
\mathbb{E}_l\left[\frac{\sum\limits_{m=0}^{N_{\text{max}}}{O}(\mathbf{r},n,m)\psi^{(k)}(\mathbf{r},m)}{\psi^{(l)}(\mathbf{r},n)}\right]}.
\end{equation}
Notice that the equivalent of Eq.~(\ref{eq:exp_O_kl}) was derived in Eq.~(15) of Ref.~\cite{Entwistle2023Electronic} for purely electronic systems, thus with $\mathbb{E}_k[\cdot]$ only representing an average over the electronic degrees of freedom, and without the sums over $m$. 
Given the formal analogy between the expressions, the calculation of the gradient of observables with respect to $\theta$ leads to the equivalent of Eq.~(16) of Ref.~\cite{Entwistle2023Electronic}, i.e.
\begin{eqnarray}
\label{eq:grad_kl_ph}
\partial_\theta\langle\mathcal{O}\rangle_{k,l}&=&
\frac{1}{\langle\mathcal{O}\rangle_{k,l}} 
\scalebox{1.8}{\Bigg\{}
\mathbb{E}_k\left[\left(   \frac{\sum\limits_{m=0}^{N_{\text{max}}}{O}(\mathbf{r},n,m)\psi^{(l)}(\mathbf{r},m)}{\psi^{(k)}(\mathbf{r},n)} -  \mathbb{E}_k\left[ \frac{\sum\limits_{m=0}^{N_{\text{max}}}{O}(\mathbf{r},n,m)\psi^{(l)}(\mathbf{r},m)}{\psi^{(k)}(\mathbf{r},n)}\right] \right)\partial_\theta\ln|\psi^{(k)}(\mathbf{r},n)|    \right] 
\nonumber\\
&&\times
\mathbb{E}_l\left[ \frac{\sum\limits_{m=0}^{N_{\text{max}}}{O}(\mathbf{r},n,m)\psi^{(k)}(\mathbf{r},m)}{\psi^{(l)}(\mathbf{r},n)}\right]
 + (k \Longleftrightarrow l) 
\scalebox{1.8}{\Bigg\}},
\end{eqnarray}
\end{widetext}
where $(k \Longleftrightarrow l)$ represents an additional term corresponding to the expression in curly brackets with exchanged indices $k$ and $l$.

\subsection{Loss function and its gradient}
Using Eqs.~(\ref{eq:exp_O_kl}) and (\ref{eq:grad_kl_ph}) we can evaluate any observable $\mathcal{O}$ and its gradient, including the loss function. In the following, we provide the explicit expression of $O(\mathbf{r},n,m)$ to evaluate the loss function and its gradient.

Following the discussion in Subsec.~\ref{subsec:vm_loss}, the Pauli-Fierz Hamiltonian consists of four terms in the following form:
\begin{eqnarray}\label{eq:Hamiltonian_all_2}
H_\text{PF}&=&H_\text{e}+V_{\text{ph}}+V_{\text{el-ph}}+V_{\text{dip}}\nonumber\\
&=&H_\text{e}+\omega \hat{b}^\dagger \hat{b}
-\sqrt{\frac{\omega}2}\lambda(\boldsymbol{\varepsilon}\cdot\mathbf{d})(\hat{b}+\hat{b}^\dagger)\nonumber\\
&&+\frac12\lambda^2(\boldsymbol{\varepsilon}\cdot\mathbf{d})^2,
\end{eqnarray}
\noindent with molecular dipole operator $\mathbf{d}=-\sum\limits_{i=1}^{N_\text{e}}\mathbf{r}_i+\sum\limits_{I=1}^{N_\text{nuc}}Z_I\mathbf{R}_I$ for electron positions $\{\mathbf{r}_i\}_{i=1}^{N_\text{e}}$ and nuclei positions $\{\mathbf{R}_I\}_{I=1}^{N_{\text{nuc}}}$. The first term $H_{\text{e}}$ is the electronic Hamiltonian
\begin{eqnarray}\label{eq:Hamiltonian_electronic_2}
H_{\text{e}}&=&-\sum\limits_{i=1}^{N_\text{e}}\frac12\nabla_i^2-\sum\limits_{i=1}^{N_\text{e}}\sum\limits_{I=1}^{N_{\text{nuc}}}\frac{Z_I}{|\mathbf{r}_i-\mathbf{R}_I|}\\
&&
+\frac12\sum\limits_{i\neq j}^{N_\text{e}}\frac1{|\mathbf{r}_i-\mathbf{r}_j|}\nonumber+\frac12\sum\limits_{I\neq J}^{N_{\text{nuc}}}\frac{Z_IZ_J}{|\mathbf{R}_I-\mathbf{R}_J|},
\end{eqnarray}
\noindent where $\nabla_i^2$ is the Laplace operator w.r.t. $\mathbf{r}_i$ and $Z_I$ is the nuclear charge number of nucleus $I$. The second term
\begin{equation}\label{eq:V_pho-ho}
V_{\text{ph}}=\omega \hat{b}^\dagger \hat{b}
\end{equation}
\noindent is the energy of harmonic oscillators of the photonic mode neglecting the zero-point energy, which is purely photonic. The electron-photon coupling term
\begin{equation}\label{eq:V_el-ph}
V_{\text{el-ph}}=-\sqrt{\frac{\omega}2}\lambda(\boldsymbol{\varepsilon}\cdot\mathbf{d})(\hat{b}+\hat{b}^\dagger)
\end{equation}
\noindent describes the coupling between the electronic and photonic degrees of freedom. The last term
\begin{equation}\label{eq:V_dip}
V_{\text{dip}}=\frac12\lambda^2(\boldsymbol{\varepsilon}\cdot\mathbf{d})^2
\end{equation}
\noindent is the dipole self-energy. Ref.~\cite{Rokaj2018Light} shows this term is indispensable because it ensures that the Hamiltonian is gauge and origin invariant and bounded from below.

From Eq.~\eqref{eq:Hamiltonian_all_2}, we have that 
\begin{eqnarray}
    H_{\text{PF}}(\mathbf{r},n,m) &=&
    H_\text{e}(\mathbf{r},n,m) + V_{\text{ph}}(\mathbf{r},n,m) \nonumber\\&&+ V_{\text{el-ph}}(\mathbf{r},n,m) + V_{\text{dip}}(\mathbf{r},n,m),
\end{eqnarray}
\noindent where
\begin{eqnarray}\label{eq:E_e}
    H_\text{e}(\mathbf{r},n,m) &=& \delta_{n,m}\left(-\sum\limits_{i=1}^{N_\text{e}}\frac12\nabla_i^2-\sum\limits_{i=1}^{N_\text{e}}\sum\limits_{I=1}^{N_{\text{nuc}}}\frac{Z_I}{|\mathbf{r}_i-\mathbf{R}_I|}\right.\nonumber\\
&&+\frac12\sum\limits_{i\neq j}^{N_\text{e}}\frac1{|\mathbf{r}_i-\mathbf{r}_j|}\left.+\frac12\sum\limits_{I\neq J}^{N_{\text{nuc}}}\frac{Z_IZ_J}{|\mathbf{R}_I-\mathbf{R}_J|}\right),\nonumber\\
\end{eqnarray}
\begin{equation}
\label{eq:E_ph}
V_\text{ph}(\mathbf{r},n,m) = \delta_{n,m}\omega n,
\end{equation}
\begin{eqnarray}\label{eq:E_el-ph}
    V_\text{el-ph}(\mathbf{r},n,m) &=&
    - \sqrt{\frac\omega2}\lambda(\boldsymbol{\varepsilon}\cdot\mathbf{d}(\mathbf{r}))\nonumber\\
    &&\times\left(\delta_{n+1,m}\sqrt{n+1}+\delta_{n-1,m}\sqrt{n}\right),
\end{eqnarray}
\begin{equation}\label{eq:E_dip}
V_\text{dip}(\mathbf{r},n,m) = \delta_{n,m}\frac12\lambda^2(\boldsymbol{\varepsilon}\cdot\mathbf{d}(\mathbf{r}))^2,
\end{equation}
$\mathbf{d}(\mathbf{r})=-\sum\limits_{i=1}^{N_\text{e}}\mathbf{r}_i+\sum\limits_{I=1}^{N_{\text{nuc}}}Z_I\mathbf{R}_I$ representing the dipole moment.

The expression for the squared spin operator $S^2(\mathbf{r},n,m)$, appearing in the loss function, can be computed as in Ref.~\cite{liu2023calculate, Szabo2024An}, yielding
\begin{eqnarray}\label{eq:exp_S2_2}
S^2(\mathbf{r},n,m)
&=&\delta_{n,m}\Biggl(-\frac{N_{\uparrow}-N_{\downarrow}}4\left(N_{\uparrow}-N_{\downarrow}+2\right)+N_{\downarrow}\nonumber\\
&&-\sum\limits_{i=1}^{N_{\uparrow}}\sum\limits_{j=N_{\uparrow}+1}^{N_\text{e}}\left[\frac{\psi(\mathbf{\tilde{r}}_{ij},n)}{\psi(\mathbf{r},n)}\right]\Biggr),
\end{eqnarray}
where $N_{\uparrow}$ and $N_{\downarrow}$ are the number of spin-up and spin-down electrons respectively, $\mathbf{\tilde{r}}_{ij}$ is obtained by exchanging the electron positions $\mathbf{r}_i$ and $\mathbf{r}_j$ in $\mathbf{r}$.

The overlap between different states, appearing in the loss function, can be computed choosing $\mathcal{O}$ as the identity, i.e. $\mathcal{O}(\mathbf{r},n,m) = \delta_{n,m}$.

\subsection{Electron density of a state}\label{app:electron_density}
The electron density $\rho_\text{e}(\vec{r})$ measures the probability of observing an electron around the point $\vec{r}\in\mathbb{R}^3$. It can be computed using Eq.~(\ref{eq:exp_O_kl}) with $O(\mathbf{r},n,m)=\rho_\text{e}(\vec{r})[\mathbf{r},n,m]$, where
\begin{equation}\label{eq:e_density_MC}
\rho_\text{e}(\vec{r})[\mathbf{r},n,m] = \delta_{n,m}
\frac1{N_{\text{e}}}\sum\limits_{i=1}^{N_\text{e}}\delta(\vec{r}-\mathbf{r}_i).
\end{equation}
In Fig.~\ref{fig:H22_shift_and_dipole} of the main text we plot the electron density along one spatial direction for each curve, by integrating over two other coordinates. For example, the electron density along $x$ axis is defined as $\rho_\text{e}(x) = \int\int \rho_\text{e}(\vec{r}) \dd y \dd z$. In practice, this is computed generating Monte Carlo samples of electron positions, and only considering one of the three coordinates of each sample. We then use kernel density estimation, replacing the delta functions in Eq.~\eqref{eq:e_density_MC} with narrow Gaussian kernels~\cite{Scott2015Multivariate}.

\subsection{Density matrix of the photonic state}\label{app:photon_dm}
Given a state $|\psi^{(k)}\rangle$, we are interested in computing the reduced density matrix describing the photonic system. This can be computed by taking the partial trace relative to the electronic degrees of freedom, i.e.
\begin{equation}
\rho_{\text{ph}}=\frac{\mathrm{Tr}_{\mathbf{r}}(|\psi^{(k)}\rangle\langle\psi^{(k)}|)}{\mathrm{Tr}(|\psi^{(k)}\rangle\langle\psi^{(k)}|)}
=\frac{\int\mathrm{d}\mathbf{r}\langle\mathbf{r}|\psi^{(k)}\rangle\langle\psi^{(k)}|\mathbf{r}\rangle}{\sum\limits_{m}\int\mathrm{d}\mathbf{r}^\prime|\psi^{(k)}(\mathbf{r}^\prime,m)|^2}.
\end{equation}
We are interested in expressing the photonic density matrix in the Fock basis, i.e. we want to compute the matrix elements
\begin{widetext}
\begin{eqnarray}
\label{eq:rho_ph_element}
\langle n_1|\rho_{\text{ph}}|n_2\rangle &=& 
\frac{\int\mathrm{d}\mathbf{r} \, \psi^{(k)}(\mathbf{r},n_1)\psi^{(k)}(\mathbf{r},n_2)}{\sum\limits_{m}\int\mathrm{d}\mathbf{r}^\prime|\psi^{(k)}(\mathbf{r}^\prime,m)|^2} =
\int\mathrm{d}\mathbf{r}\sum\limits_{n} \frac{|\psi^{(k)}(\mathbf{r},n)|^2}{{{\sum\limits_{m}\int\mathrm{d}\mathbf{r}^\prime|\psi^{(k)}(\mathbf{r}^\prime,m)|^2}}}\times\sum\limits_{m}  \delta_{n,n_1} \delta_{m,n_2}  \frac{\psi^{(k)}(\mathbf{r},m)}{\psi^{(k)}(\mathbf{r},n)}\nonumber\\
&=&\mathbb{E}_k\left[ \sum\limits_{m}  \delta_{n,n_1} \delta_{m,n_2}  \frac{\psi^{(k)}(\mathbf{r},m)}{\psi^{(k)}(\mathbf{r},n)} \right],
\end{eqnarray}
\end{widetext}
where we used that $\psi^{(k)}(\mathbf{r},n)$ is real. Comparing with Eq.~(\ref{eq:exp_O_kl}), we notice that Eq.~(\ref{eq:rho_ph_element}) can be seen as the Monte Carlo expectation value of the operator $O(\mathbf{r},n,m)$
\begin{equation}
    O(\mathbf{r},n,m) = \delta_{n,n_1}\delta_{m,n_2}
\end{equation}
over the $|\psi^{(k)}(\mathbf{r},n)\rangle$ state. Since $|\psi^{(k)}(\mathbf{r},n)\rangle$ is real and Hermitian, $\langle n_1|\rho_{\text{ph}}|n_2\rangle=\langle n_2|\rho_{\text{ph}}|n_1\rangle$. This can be used to obtain the real and Hermitian operator
\begin{equation}
    O(\mathbf{r},n,m) = \frac{1}{2}\left(\delta_{n,n_1}\delta_{m,n_2} + \delta_{m,n_1}\delta_{n,n_2}\right)
\end{equation}
whose expectation value $\ev{\mathcal{O}}_{k,k}$, computed with Eq.~(\ref{eq:exp_O_kl}), yields the photonic density matrix element $\langle n_1|\rho_{\text{ph}}|n_2\rangle$.

\section{Sampling algorithm}\label{app:sampling}
In order to compute the expectation values $\mathbb{E}_k[\cdot]$ using Monte Carlo, we need to draw samples of $(\mathbf{r},n)$ distributed according to the probability distribution $P(\mathbf{r},n)\propto|\psi^{(k)}(\mathbf{r},n)|^2$. This can be done using a generalization of the Metropolis-Hastings algorithm to discrete and continuous samples~\cite{Metropolis1953Equation,Hastings1970Monte}. In this algorithm, we generate a reversible Markov chain in state space $\mathbb{R}^{3N_{\text{e}}}\times\{0,1,\cdots,N_{\text{max}}\}$, which has a stationary distribution $P(\mathbf{r},n)\propto|\psi^{(k)}(\mathbf{r},n)|^2$~\cite{Robert2004Monte}. It consists of two parts: proposal and acceptance-rejection. Given a sample $(\mathbf{r},n)$, we propose a new sample $(\mathbf{r}',n')$ in state space according to the conditional probability $g((\mathbf{r}',n')\mid(\mathbf{r},n))$, and accept it with probability $A((\mathbf{r}',n'),(\mathbf{r},n))$. The overall transition probability from state $(\mathbf{r},n)$ to $(\mathbf{r}',n')$ is then
\begin{equation}
P((\mathbf{r}',n')\mid(\mathbf{r},n))=g((\mathbf{r}',n')\mid(\mathbf{r},n))\times A((\mathbf{r}',n'),(\mathbf{r},n)).
\end{equation}
\noindent The detailed balance condition for the Markov chain reads
\begin{equation}
P((\mathbf{r}',n')\mid(\mathbf{r},n))\times P(\mathbf{r},n)=P((\mathbf{r},n)\mid(\mathbf{r}',n'))\times P(\mathbf{r}',n'),
\end{equation}
\noindent and it is satisfied when we set the acceptance probability to
\begin{eqnarray}
 A((\mathbf{r}',n'),(\mathbf{r},n))
= \min\left\{\frac{P(\mathbf{r}',n')}{P(\mathbf{r},n)}\times\frac{g((\mathbf{r},n)\mid(\mathbf{r}',n'))}{g((\mathbf{r}',n')\mid(\mathbf{r},n))},1\right\}
\nonumber\\
=\min\left\{\frac{|\psi^{(k)}(\mathbf{r}',n')|^2}{|\psi^{(k)}(\mathbf{r},n)|^2}\times\frac{g((\mathbf{r},n)\mid(\mathbf{r}',n'))}{g((\mathbf{r}',n')\mid(\mathbf{r},n))},1\right\}.\nonumber
\end{eqnarray}
We first initialize a set of random walkers in the state space. For each walker, we evaluate the wavefunction $\psi^{(k)}(\mathbf{r},n)$ in the current state $(\mathbf{r},n)$. Then we propose a new sample $(\mathbf{r}',n')$ according to $g((\mathbf{r}',n')\mid(\mathbf{r},n))$ and evaluate $\psi^{(k)}(\mathbf{r}',n')$. We then accept the new sample with probability $ A((\mathbf{r}',n'),(\mathbf{r},n))$.
In order to sample from the joint distribution of continuous ($\mathbf{r}$) and discrete ($n$) variables, we introduce the following conditional probability:
\begin{equation}\label{eq:transition_p}
\begin{aligned}
g((\mathbf{r}',n')\mid(\mathbf{r},n))=&\frac12\delta_{n,n'}g_r(\mathbf{r}'\mid \mathbf{r})
+\frac12\delta\left(\mathbf{r}-\mathbf{r}'\right)g_n(n'\mid n),
\end{aligned}
\end{equation}
\noindent where $g_r(\mathbf{r}'\mid \mathbf{r})$ and $g_n(n'\mid n)$ are respectively a continuous and discrete conditional distribution. In practice, this amounts to the following:
with probability $50\%$, we fix $n$ and propose a change to $\mathbf{r}$ according to $g_r(\mathbf{r}'\mid \mathbf{r})$; with probability $50\%$, we propose a change to $n$ using $g_n(n'\mid n)$ with $\mathbf{r}$ fixed. 

For the continuous proposal function $g_r(\mathbf{r}'\mid \mathbf{r})$, we use an adjustable normal distribution $\mathbf{r}'\sim\mathcal{N}(\mathbf{r},\tau^2)$ with tunable parameter $\tau$. The idea of using a tunable probability distribution for the proposal is to adjust the acceptance rate of all the walkers to be as close as possible to the rate of 0.57~\cite{Gelman1997Weak,Roberts2001Optimal}. If the acceptance rate of the samples where $\mathbf{r}$ is updated is larger than 0.57, we decrease it by increasing $\tau$, the ``exploring radius". Thus the proposals $\mathbf{r}$ are less likely to be accepted and the theoretical acceptance rate is lowered. On the contrary, if the acceptance rate is too small, we decrease $\tau$ for the next step. Indeed, in the limit of zero sampling radius, the acceptance ratio is 1. See Alg.~\ref{alg:2-stepMH} for details.

For the discrete proposal function $g_n(n'\mid n)$, we sample from the probability distribution
\begin{equation}\label{eq:Delta_n}
\left\{\begin{aligned}
&\frac{p^{|n'-n|}}{1+2p+2p^2},\quad&\text{if }|n'-n|\leq2,\\
&0,\quad&\text{otherwise},
\end{aligned}\right.
\end{equation}
\noindent and clip the value of $n'$ into the range of $\{0,1,\cdots,N_{\text{max}}\}$. Here, $p$ is a tunable parameter. Notice that, because of the clipping, $g_n(n'\mid n)$ is no longer symmetric regarding to $n$ and $n'$, i.e.
\begin{equation}
\frac{g_n(1\mid0)}{g_n(0\mid1)}=\frac{g_n(N_{\text{max}}-1\mid N_{\text{max}})}{g_n(N_{\text{max}}\mid N_{\text{max}}-1)}=\frac1{1+p},
\end{equation}
\noindent and
\begin{equation}
\frac{g_n(n\mid n')}{g_n(n'\mid n)}=1
\end{equation}
\noindent in all other cases.
Similar to the continuous case, if the acceptance rate of the samples where $n$ is updated is larger than 0.57, we decrease it by increasing the variance $\sigma$ of the discrete distribution, which plays the role of the ``exploring radius" in the discrete case. On the contrary, if the acceptance rate is too small, we decrease $\sigma$ for the next step. See Alg.~\ref{alg:2-stepMH} for details.
In order to tune the variance $\sigma$, we update the parameter $p$. The relation between $p$ and $\sigma$ can be computed exactly ignoring the clipping of $n^\prime$, and it is given by
\begin{equation}
\sigma^2=\frac{2(p+4p^2)}{1+2p+2p^2},
\end{equation}
\noindent allowing us to express $p$ as a function of $\sigma$:
\begin{equation}\label{eq:p(sigma)}
p(\sigma)=\frac{\sigma^2-1+\sqrt{-\sigma^4+6\sigma^2+1}}{8-2\sigma^2}.
\end{equation}

The overall algorithm is shown in Alg.~\ref{alg:2-stepMH}.
\begin{algorithm}
\caption{Discrete-continuous Metropolis-Hastings}\label{alg:2-stepMH}
\KwData{wavefunction $\psi^{(k)}:\mathbb{R}^{3N_\text{e}}\times\{0,1,\cdots,N_{\text{max}}\}\rightarrow\mathbb{R}$, $(\mathbf{r},n)\mapsto\psi^{(k)}(\mathbf{r},n)$, number of sampling steps $N_{\text{steps}}\in\mathbb{N}_+$, number of walkers $N_\text{w}$}
\KwResult{samples $\{(\mathbf{r}_t^{(w)},n_t^{(w)})\mid t\in[N_{\text{steps}}],w\in[N_\text{w}]\}$ asymptotically with density
$\propto|\psi^{(k)}(\mathbf{r},n)|^2$}
set $t\gets0$\;
set $\tau_{r,0}\gets1$\;
set $\tau_{n,0}\gets\sqrt{2}$\;
initialize $\mathbf{r}_0$, $n_0$\;
\While{$t<N_{\mathrm{steps}}$}{
    flip a fair coin (head or tail) $N_\mathrm{w}$ times independently\;
    \For{$\mathrm{walker}$ $w$ $\mathrm{with}$ $\mathrm{result}$ $\mathrm{head}$}{
    set $n_{t+1}^{(w)}\gets n_{t}^{(w)}$\;
    sample $\mathbf{r}'$ according to normal distribution $\mathcal{N}(\mathbf{r}_t^{(w)},\tau_{r,t}^2)$ and
    propose $\mathbf{r}'$\;
    accept and set $\mathbf{r}_{t+1}^{(w)}\gets\mathbf{r}'$ w.p. $\min\left\{\frac{|\psi^{(k)}(\mathbf{r}',n_t^{(w)})|^2}{|\psi^{(k)}(\mathbf{r}_t^{(w)},n_t^{(w)})|^2},1\right\}$, set $\mathbf{r}_{t+1}^{(w)}\gets\mathbf{r}_t^{(w)}$ otherwise\;}
    calculate the acceptance rate $\alpha_r$ among all walkers with result head\;
    set $\tau_{r,t+1}\gets\tau_{r,t}\times\frac{\max\{\alpha_r,0.05\}}{0.57}$
    set $p\gets\frac{\tau_{n,t}^2-1+\sqrt{-\tau_{n,t}^4+6\tau_{n,t}^2+1}}{8-2\tau_{n,t}^2}$\;
    \For{$\mathrm{walker}$ $w$ $\mathrm{with}$ $\mathrm{result}$ $\mathrm{tail}$}{
    set $\mathbf{r}_{t+1}^{(w)}\gets \mathbf{r}_{t}^{(w)}$\;
    sample $n'$ from $\{n-2, n-1, n, n+1, n+2\}$ w.p. $\left\{\frac{p^2}{1+2p+2p^2},\frac{p}{1+2p+2p^2},\frac{1}{1+2p+2p^2},\frac{p}{1+2p+2p^2},\frac{p^2}{1+2p+2p^2}\right\}$ respectively\;
    set $n'\gets \max\{0,\min\{N_{\text{max}},n'\}\}$ and propose $n'$\;
    accept and set $n_{t+1}^{(w)}\gets n'$ w.p. $\min\left\{\frac{|\psi^{(k)}(\mathbf{r}_{t+1}^{(w)},n')|^2}{|\psi^{(k)}(\mathbf{r}_{t+1}^{(w)},n_t^{(w)})|^2}\times\frac{g_n(n_t^{(w)}\mid n')}{g_n(n'\mid n_t^{(w)})},1\right\}$, set $n_{t+1}^{(w)}\gets n_t^{(w)}$ otherwise\;}
    calculate the acceptance rate $\alpha_n$ among all walkers with result tail\;
    set $\tau_{n,t+1}\gets\tau_{n,t}\times\frac{\max\{\alpha_n,0.05\}}{0.57}$\;
    update $\psi^{(k)}$ in the training process\;
}
\end{algorithm}

\section{Squeezing of electromagnetic field}\label{app:squeezing}

To derive the effective Hamiltonian for the photon field by eliminating the electronic degrees of freedom, we begin with the Pauli-Fierz Hamiltonian in Eq.~\eqref{eq:Hamiltonian_all}, expressed as 
\begin{equation}
  H_\text{PF} = \tilde{H}_\text{e} + \omega \hat{b}^\dagger \hat{b} + V_{\text{el-ph}},  
\end{equation}
\noindent where \( \tilde{H}_\text{e} = H_\text{e} + V_{\text{dip}} \) is the electronic Hamiltonian renormalized by the self-interaction, \( \omega \hat{b}^\dagger \hat{b} \) represents the photon Hamiltonian, and the electron-photon coupling term is given by
\begin{equation}
   V_{\text{el-ph}}= - \sqrt{\frac{\omega}{2}} \lambda (\varepsilon \cdot \mathbf{d})(\hat{b} + \hat{b}^\dagger) . 
\end{equation}
First of all, we expand \( \tilde{H}_\text{e} \) in terms of its electronic eigenstates \( |\tilde{\psi}^{(M)}_{\rm e}\rangle \), which satisfy \( \tilde{H}_\text{e} |\tilde{\psi}^{(M)}_{\rm e}\rangle = \tilde{E}_M |\tilde{\psi}^{(M)}_{\rm e}\rangle \), with \( M = 0, 1, \dots \). Here, \( \tilde{E}_0 \) is the ground state energy.
Assuming \( V_{\text{el-ph}} \) is a small perturbation, we split the total Hamiltonian into an unperturbed part, \( H_0 = \tilde{H}_\text{e} + \omega \hat{b}^\dagger\hat{b} \), and a perturbation, \( V_{\text{el-ph}} \). Using second-order perturbation theory, we trace out the electronic degrees of freedom to obtain an effective Hamiltonian for the photon field.

The second-order correction to the effective photonic Hamiltonian is thus:
\begin{equation}
\delta H =- \sum_{M \neq 0} \frac{|\langle \tilde{\psi}^{(0)}_{\rm e} | V | \tilde{\psi}^{(M)}_{\rm e}  \rangle|^2}{\tilde{E}_M -\tilde{E}_0}=-\chi(\hat{b} + \hat{b}^\dagger)^2,
\end{equation}
\noindent where
\begin{equation}
\chi = \frac{\omega \lambda^2}{2} \sum_{M \neq 0} \frac{|\boldsymbol{\varepsilon} \cdot \langle \tilde{\psi}^{(0)}_{\rm e} | \mathbf{d} | \tilde{\psi}^{(M)}_{\rm e} \rangle|^2}{\tilde{E}_M - \tilde{E}_0}.
\end{equation}
This quantity is positive as it involves squared dipole matrix elements and positive energy denominators. By expanding the photon operators as \((\hat{b} + \hat{b}^\dagger)^2 = \hat{b}^2 + \hat{b}^{\dagger 2} + 2 \hat{b}^\dagger \hat{b} + 1\) we can write the total effective photonic Hamiltonian $H_{\rm eff}= \omega \hat{b}^\dagger\hat{b}+\delta H$ as 
\begin{equation}
H_\text{eff} = (\omega - 2\chi) \hat{b}^\dagger \hat{b} - \chi \left( \hat{b}^2 + \hat{b}^{\dagger 2} \right) - \chi.
\end{equation}
The ground state of the effective Hamiltonian \( H_\text{eff} \) is thus a squeezed vacuum state.  To determine this state, the effective Hamiltonian can be diagonalized through a Bogoliubov transformation. In this approach, we define new bosonic operators \( \hat{c} \) and \( \hat{c}^\dagger \) as linear combinations of the original photon operators, such that the Hamiltonian takes a diagonal form, as:
\begin{equation}
 \hat{b} =  \hat{c} \cosh r +  \hat{c}^\dagger \sinh r,
\end{equation}
\noindent where \( r \) is the squeezing parameter to be determined. 
Substituting this transformation into the Hamiltonian, we eliminate the non-diagonal terms involving \(  \hat{c}^2 \) and \(  \hat{c}^{\dagger 2} \), choosing the squeezing parameter \( r \) such that:
\begin{equation}
\label{eq:r}
\tanh (2 r) = -\frac{2 \chi}{ \omega - 2\chi}.
\end{equation}

In the perturbative regime, where \( \omega > 2\chi \), this equation implies that \( r \) is a real and negative quantity.
With this choice of \( r \), the Hamiltonian simplifies to:
\begin{equation}
H_\text{eff} = \hbar \tilde{\omega} \left(  \hat{c}^\dagger  \hat{c} + \frac{1}{2} \right) + E'_\text{shift},
\end{equation}
\noindent where the new frequency \( \tilde{\omega} \) is given by:
\begin{equation}
\tilde{\omega} = \sqrt{ \omega(\omega-4\chi) },
\end{equation}
\noindent and \( E'_\text{shift} \) includes all constant terms resulting from the transformation. The ground state of the effective Hamiltonian is thus the vacuum state \( | 0_c \rangle \) of the \(\hat{c} \) operators, which is related to the original photonic ground state \( | 0\rangle \) by the squeezing transformation:
\begin{equation}
| 0_c \rangle = S(r) | 0 \rangle,
\end{equation}
\noindent where \( S(r) \) is the squeezing operator defined as:
\begin{equation}
S(r) = \exp \left[ \frac{1}{2} r \left( \hat{b}^2 - \hat{b}^{\dagger 2} \right) \right].
\end{equation}
Therefore, the ground state of \( H_\text{eff} \) is a squeezed vacuum state characterized by the squeezing parameter \( r \), reflecting the modifications to the photon field due to the interaction with the electronic degrees of freedom.

In the squeezed vacuum state \( |0_c \rangle \), the variances of \( \hat{x} \) and \(\hat{p} \) are given by:
\begin{equation}
{\rm Var}[\hat{x}] = \langle 0_c | \hat{x}^2 | 0_c \rangle - \langle 0_c | \hat{x} | 0_c \rangle^2,
\end{equation}
\begin{equation}
{\rm Var}[\hat{p}] = \langle 0_c | \hat{p}^2 | 0_c \rangle - \langle 0_c | \hat{p} | 0_c \rangle^2.
\end{equation}

For the squeezed vacuum state, these variances become:
\begin{equation}
{\rm Var}[\hat{x}] = \frac{e^{-2r}}{2}, \quad {\rm Var}[\hat{p}] = \frac{e^{2r}}{2}.
\end{equation}
This indicates that the $\hat{x}$-quadrature variance is increased by a factor of \( e^{-2r} \) as $r$ is a negative number (as evident from Eq.~\eqref{eq:r}) while the  $\hat{p}$-quadrature variance is increased by \( e^{2r} \). 
This behavior aligns with the shape of the Wigner function shown in Fig.~\ref{fig:wigner_x}, where the squeezing effect manifests as an elliptical deformation, with one axis contracted and the other elongated, corresponding to the respective quadrature variances.

\section{Density matrix of the photonic wavefunction}\label{app:wigner}
Here we present explicitly the photonic density matrices discussed in Sec.~\ref{sec:photon_wf} and Sec.~\ref{sec:photon_wf_e}.

\begin{widetext}
\begin{itemize}
\item At $R=4.0\text{ \AA}$, the photonic density matrix of the ground state of two H\textsubscript{2} in the cavity with polarization $\boldsymbol{\varepsilon}_x$
, denoted as $\rho_{\text{ph}}^{\boldsymbol{\varepsilon}_x}$, is shown in Eq.~\eqref{eq:photon_dm_x}.
The diagonal terms for $n=0$ and $1$ are bold to show the zero-photon coefficient is the most dominant. The state is very close to the ground Fock state $|0\rangle$. 
\begin{equation}\label{eq:photon_dm_x}
\rho_{\text{ph}}^{\boldsymbol{\varepsilon}_x}=\bordermatrix{n & 0 & 1 & 2 & 3 & 4 & 5\cr
0 & \textbf{992.621} & -0.079 & 11.078 & 0.761 & 0.137 & 0.090 \cr
1 & -0.079 & \textbf{7.181} & 0.027 & 0.070 & 0.011 & -0.004 \cr
2 & 11.078 & 0.027 & 0.195 & 0.017 & 0.003 & 0.002 \cr
3 & 0.761 & 0.070 & 0.017 & 0.002 & 0.000 & 0.000 \cr
4 & 0.137 & 0.011 & 0.003 & 0.000 & 0.000 & 0.000 \cr
5 & 0.090 & -0.004 & 0.002 & 0.000 & 0.000 & 0.000
}\times10^{-3},
\end{equation}

\item The photonic density matrix of the first singlet excited state of H\textsubscript{2} in the cavity with polarization parallel to its main axis, denoted as $\rho_{\text{ph}}$, is shown in Eq.~\eqref{eq:photon_dm_e}. The state is close to a superposition of the $|0\rangle$ and $|1\rangle$ photonic states, with the first two diagonal entries being $\simeq\frac12$ (boldfaced).
\begin{equation}\label{eq:photon_dm_e}
\rho_{\text{ph}}=\bordermatrix{n & 0 & 1 & 2 & 3 & 4 & 5\cr
0 & \textbf{445.762} & 1.949 & 26.585 & 0.621 & 0.174 & 0.306 \cr
1 & 1.949 & \textbf{552.526} & 2.279 & 1.081 & 0.290 & 0.343 \cr
2 & 26.585 & 2.279 & 1.704 & 0.043 & 0.012 & 0.021 \cr
3 & 0.621 & 1.081 & 0.043 & 0.006 & 0.001 & 0.003 \cr
4 & 0.174 & 0.290 & 0.012 & 0.001 & 0.000 & 0.001 \cr
5 & 0.306 & 0.343 & 0.021 & 0.003 & 0.001 & 0.002}\times10^{-3}.
\end{equation}
\end{itemize}

\section{CASSCF calculations for H\textsubscript{2}}\label{app:CASSCF_H2}
\begin{figure*}[htb!]
\centering
\includegraphics[width=17.2cm]{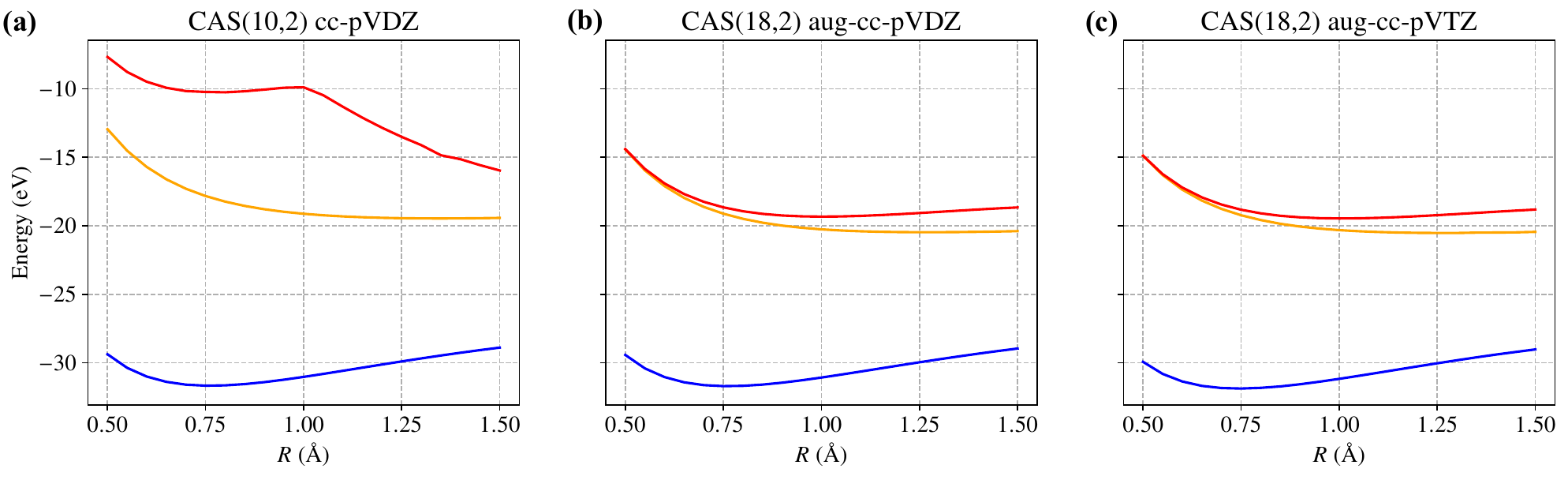}
\caption{\textbf{Energies of H\textsubscript{2} singlet states with various bond length.} We use CASSCF with two electrons and different basis sets to calculate the first three lowest energies. \textbf{(a)}: cc-pVDZ basis with 10 orbitals (FCI), \textbf{(b)}: aug-cc-pVDZ basis with 18 orbitals (FCI), \textbf{(c)}: aug-cc-pVTZ basis with 18 orbitals. The more accurate results match those obtained using our proposed deep QMC method.} 
\label{fig:H2_CASSCF}
\medskip
\small
\end{figure*}
\end{widetext}

In this section, we present the three lowest energy singlet states of H\textsubscript{2}, calculated using the complete active space self-consistent field (CASSCF) method~\cite{Siegbahn1980A,Roos1980A,Siegbahn1981The} with the PySCF package~\cite{Sun2020Recent}. We investigate how increasing the size of the basis set, from cc-pVDZ to aug-cc-pVDZ and aug-cc-PVTZ, affects the energy gap between the first and second excited states. Specifically, for cc-pVDZ, we perform a CAS(10,2) calculation [Fig.~\ref{fig:H2_CASSCF}(a)], selecting both electrons and all ten orbitals as the active space, which is equivalent to Full Configuration Interaction (FCI). For the other basis sets, we use a CAS(18,2) calculation [Fig.~\ref{fig:H2_CASSCF}(b) and (c)], where for aug-cc-pVDZ, this also corresponds to FCI. We find that the results for cc-pVDZ are in good agreement with reference~\cite{Haugland2020Coupled}, while the more accurate calculations using aug-cc-pVDZ and aug-cc-pVTZ, which display a significantly smaller energy gap between the first (orange) and second (red) excited states, are consistent with our findings [see Fig.~\ref{fig:H2_energy}].

\section{Hyperparameters}\label{app:hyperparameters}
Here we list the hyperparameters used in generating the data presented in Sec.~\ref{sec:results} as Tab.~\ref{tab:hyperparameters}.

\begin{table}[h!]
\centering
\begin{tabular}{c c} 
\hline\hline
Hyperparameter & Value \\ [0.5ex] 
\hline
Cavity frequency $\omega$ for two H\textsubscript{2} [cf. Eq.~\eqref{eq:Hamiltonian_all}] & 12.7 eV \\
Cavity frequency $\omega$ for H\textsubscript{2} [cf. Eq.~\eqref{eq:Hamiltonian_all}] & 12.750702 eV \\
Coupling strength $\lambda$ for two H\textsubscript{2} [cf. Eq.~\eqref{eq:Hamiltonian_all}] & 0.1 \\
Coupling strength $\lambda$ for H\textsubscript{2} [cf. Eq.~\eqref{eq:Hamiltonian_all}] & 0.05 \\
Maximum number of determinants & 16 \\
Embedding dimension & 128 \\
Number of interaction layers & 3 \\
Number of hidden layers in $w_\theta$~\cite{Schaetzle2023Deep} & 2 \\
Number of hidden layers in $h_\theta$~\cite{Schaetzle2023Deep} & 2 \\
Number of hidden layers in $\kappa_\theta$~\cite{Schaetzle2023Deep} & 2 \\
Number of hidden layers in $g_\theta$~\cite{Schaetzle2023Deep} & 1 \\
Number of hidden layers in $\eta_\theta$~\cite{Schaetzle2023Deep} & 1 \\
Photon number cutoff $N_{\text{max}}$ & 5 \\
Number of walkers & 2048 \\
Number of pretraining steps for two H\textsubscript{2} & 5000 \\
Pretraining basis for two H\textsubscript{2} & aug-cc-pVTZ \\
Number of orbitals in two H\textsubscript{2} CASSCF pretraining  & 16 \\
Number of training steps for two H\textsubscript{2} & 200000 \\
Number of evaluation steps for two H\textsubscript{2} & 50000 \\
Number of training steps for H\textsubscript{2} & 60000 \\
Number of evaluation steps for H\textsubscript{2} & 50000 \\
$\alpha$ for two H\textsubscript{2} [cf. Eq.~\eqref{eq:loss_function_ext}] & 8.0 \\
$\alpha$ for H\textsubscript{2} [cf. Eq.~\eqref{eq:loss_function_ext}] & 4.0 \\
Number of decorrelation sampling steps & 30 \\
Target acceptance rate & 57\% \\ 
Optimizer & K-FAC~\cite{Martens2015Optimizing} \\ [1ex] 
\hline\hline
\end{tabular}
\caption{Hyperparameters used in calculations.}
\label{tab:hyperparameters}
\end{table}

\end{appendices}

\clearpage
\bibliography{deepqmc_photon_excited}

\end{document}